\documentclass[preprint,showpacs,prb]{revtex4}
\include{graphicx}

\begin{document}

\title{Slow Relaxation and Equilibrium Dynamics in a 2 D Coulomb Glass: Demonstration of Stretched Exponential Energy Correlations}

\author{M. Kirkengen}\email{mki@fys.uio.no} \affiliation{Department of Physics, University
of Oslo,  P. O. Box 1048 Blindern, 0316 Oslo, Norway}
\author{J. Bergli} \email{jbergli@fys.uio.no}\affiliation{Department of Physics, University of
Oslo,  P. O. Box 1048 Blindern, 0316 Oslo, Norway} 
\date{\today}

\begin{abstract}
  We have simulated energy relaxation and equilibrium dynamics in
  Coulomb Glasses using the random energy lattice model. We show that
  in a temperature range where the Coulomb Gap is already well
  developed, ($T=0.03-0.1$) the system still relaxes to an equilibrium
  behavior within the simulation time scale. For all temperatures
  $T$, the relaxation is slower than exponential. Analyzing the energy
  correlations of the system at equilibrium $C(\tau)$, we find a
  stretched exponential behavior,
  $C(\tau)=e^{-(\tau/\tau_0)^\gamma}$. We study the temperature
  dependence of $\tau_0$ and $\gamma$. $\tau_0$ is shown to increase
  faster than exponentially with decreasing $T$. $\gamma$ is
  proportional to $T$ at low temperature, and approaches unity for
  high temperature.  We define a time $\tau_{\gamma}$ from these
  stretched exponential correlations, and show that this time
  corresponds well with the time required to reach equilibrium. From
  our data it is not possible to determine whether $\tau_{\gamma}$
  diverges at any finite temperature, indicating a glass transition,
  or whether this divergence happens at zero temperature.
  While the time dependence of the system
  energy can be well fitted by a random walker in a harmonic potential
  for high temperatures ($T=10$), this simple model fails to describe
  the long time scales observed at lower temperatures. Instead we
  present an interpretation of the configuration space as a structure
  with fractal properties, and the time evolution as a random walk on
  this fractal-like structure.
\end{abstract}

\maketitle

\section{Introduction}
In a doped semiconductor at low temperature, electron transport can be
dominated by variable range hopping of localized electrons.\cite{Mott}
If the active impurities are sufficiently close, the Coulomb
interaction between the localized electrons will give an important
contribution to the disorder of the energy levels.  Such a system is
known as a Coulomb Glass, where the term 'glass' is due to the many
features it shares with other glasses, including slow dynamics and
aging, and short distance correlations combined with long range
disorder.

Coulomb Glasses have been studied extensively both numerically and
analytically. Among the top theoretical results describing the model
is the derivation by Efros and Shklovskii of the 'Coulomb Gap', a gap
in the distribution of single electron state energies at any given
time.\cite{EfrosShklovskii}

Among the most interesting experimental results we would like to
mention the demonstration of memory effects in the conductivity
experiments performed by Ovadyahu et al.,\cite{Ovadyahu} and the large
$1/f$-noise increase related to the metal-insulator transition
observed by Kar et al.\cite{Kar}

The first simulations trying to model the Coulomb Glass were performed
by Kurosawa.\cite{Kurosawa} Later several groups have studied various
parameter spaces, using very different approaches. In all approaches
we have found, the defects have been modeled as static sites, with
the electrons jumping from site to site. The number of electrons is
usually chosen to be half that of the sites, reflecting a symmetry in
states over and under the Fermi level.

The disorder of the system can be modeled in two different ways.
First, one can choose to place the sites the electrons can jump
between in a regular lattice, and introduce disorder through a local
site energy. This means distances are functions of site index
difference only, greatly reducing calculational effort. The second
possibility is having sites of equal local site energy, and
introducing disorder through the position of the sites. This gives the
advantage of a symmetry in occupation numbers, changing the occupation
of all states gives a state of the same energy, allowing a convenient
definition of an order parameter.\cite{Davies} The applicability of
this order parameter has recently been disputed by Matulewski et
al.,\cite{MATULEWSKI} who claim that the long range interaction of the
electrons leads to a system size dependence of the order parameter,
and that the temperature where it goes to zero depends on the minimal
inter-site distance allowed. It can be argued that positional disorder
requires smaller system sizes for good statistics at low temperatures,
as the energy disorder will lead to some sites ``freezing out'' and
having fixed occupation probability.  However, it is possible that the
two models can give very different behavior, so the
further study of both is definitely justified.

The most realistic model would be to study a 3D model. Still, many
simulations, including ours, use a 2D model only. The advantage of
this is that we can study systems of much larger linear dimension,
which we assume to be important due to the long range nature of the
Coulomb interaction.

Next, choice of temperature range greatly affects the choice of
algorithm. For small systems at extremely low temperatures, it is
possible to imagine that all configurations relevant for the dynamics
can be mapped, and transitions between these
calculated.\cite{DiazSanchez,Pollak} This allows studying the effect
of transitions involving multiple simultaneous electron jumps.
Increases in size and temperature rapidly makes this method
impossible, even with the best future supercomputers. Fortunately,
increasing temperature also reduces the role of multi-electron
jumps,\cite{Pollak} as a wider range of states is accessible.

At higher temperatures, Monte Carlo simulations have been performed in
various ways, which can be divided in two groups. At high
temperatures, the Metropolis algorithm of picking a possible jump and
accepting it with a given probability is very
efficient.\cite{Metropolis,Tsigankov,Kolton,GrannanYu} At lower
temperatures, where only a small number of jumps are probable, one can
calculate all possible jump rates, and accept one jump based on
relative probabilities.\cite{MobiusThomas,Ortuno,Somoza} Various
optimizations and hybrids of these methods have also been used. Our
simulations belong to the latter category, and includes some
optimizations we have not seen elsewhere.

Our long term goal in starting this work is to study conductance in
Coulomb Glasses, and in particular to try to demonstrate the
interesting effects observed in experiments.\cite{Ovadyahu,Kar} In
order to achieve this, we need to establish which model and parameter
choice is appropriate. For example, Tsigankov et. al.\cite{Tsigankov}
argue that the variations in conductivity observed in Ovadyahu's
experiments can be reproduced for the random position model, but not
for the lattice model, both at a temperature of $T=0.04$. Our results
indicate that this may be due to the transition temperature for this
kind of behavior being lower than $T=0.04$ for the lattice model,
while it happens at a higher temperature for the position
disorder. This demonstrates how insufficient charting of the parameter
space may lead to wasted effort later. After surveying the
literature, we have realized that there is not even agreement on
whether there is a glass transition at any finite temperature for the
lattice model in 2 dimensions.

In principle, we need a complete understanding of the roles of all the
adjustable parameters; temperature, electron localization length,
disorder strength, disorder type, time scale and system size. In the
present paper, we have chosen to focus on the time scale of
equilibration of the system. In this way give an upper temperature
limit for when glassy effects can be observed, and estimate how the
time scales of these effects change with temperature.

In the analysis of the data, we have been inspired by work on spin
glasses, which show many of the same features as Coulomb glasses. This
is a field which has been studied in much greater detail. We have
based our analysis on work by Ogielski,\cite{Ogielski} whose methods
have been a powerful tool in understanding our results.  We find that
there are clear similarities in our results and those obtained for
spin glasses. For example, we have found that the correlation function
of the energy follows a stretched exponential behavior at long times,
a result we have not seen in previous discussions on Coulomb Glasses.

However, neither the distance dependent interaction nor the charge
conservation requirements found in Coulomb glasses have obvious
parallels in the spin glass. This is reflected in the fact that the
exponent we observe in the stretched exponential behaves differently
from what has been reported for spin glasses.

\section{Model}
We have used the standard 2D lattice model with
Hamiltonian 
\begin{equation}
\mathcal{H}=\sum_i U_i (n_i-\bar{n}) +
\frac{1}{2}\sum_{i,j}\frac{(n_i-\bar{n})(n_j-\bar{n})}{r_{i,j}}.
\end{equation}
Here $\bar{n}$ is the average occupancy, $n_i=0,1$ is the occupancy of
site $i$, and $r_{i,j}$ is the distance between sites $i$ and
$j$. $U_i$ is the site occupancy energy of site $i$, drawn from a
uniform distribution on the interval $[-U,U]$. 
All energies and temperatures are measured in 
units of the nearest neighbor Coulomb interaction 
$e^2/a$, where $a$ is the lattice constant and $e$ is the elementary charge. 

We use periodic boundary conditions in both dimensions, and side
lengths of $L$, forcing us to cut off the Coulomb interaction at the
distance $L/2$. $N=L^2$ is the total number of sites. In all
simulations presented here, we have used $U=1$, as this is standard in
the literature. However, it should be noted that at this value of the
disorder, the Coulomb Gap does not have the universal shape predicted
by Efros and Shklovskii,\cite{EfrosShklovskii} as shown in refs.
\onlinecite{MobiusRichterDrittler,Pikus,Glatz}. While this may be of importance
for details in the temperature dependence of various quantities, we do
not believe that it will significantly affect our results.

In order to calculate changes in the system energy $E$, we use the
``single electron energy'' $\epsilon_i$, defined by Efros and
Shklovskii, the energy gained by adding an electron to an empty site,
or required to remove the electron from an occupied site. It is given
by
\begin{equation}
\epsilon_i = U_i + \sum_{j}\frac{(n_j-\bar{n})}{r_{i,j}}.
\end{equation}
The change in system energy due to an electron hopping from site $i$
to site $j$ is then 
\begin{equation}
\Delta E_{i\rightarrow j} = \epsilon_j-\epsilon_i-1/r_{i,j}.
\end{equation}
The impurities are modeled as hydrogen-like states centered on the
lattice sites, with a localization length $a_l$ chosen to be $2/3 a$.

The relaxation has been performed by a modified Monte-Carlo simulation
optimized for low temperatures, but still capable of handling high
temperatures. 
A detailed description of the algorithm is
given in appendix \ref{algorithm}, along with a discussion of the
chosen parameters.

We follow Efros and Shklovskii\cite{EfrosShklovskii} and write the electron jump rate for the jump from the occupied site $i$ to the unoccupied site $j$ as
\begin{equation}
\Gamma_{i\rightarrow j} = t_0^{-1}e^{-2r_{i,j}/a_l}\frac{|\Delta
  E_{i \rightarrow j}|}{T_0}f(\Delta E_{i \rightarrow j}),
\label{tunnelingrate}
\end{equation}
where for processes involving phonon emission,
\begin{equation}
f(\Delta E_{i \rightarrow j}) = 1+\frac{1}{e^{|\Delta E_{i
      \rightarrow j}|/T}-1}, \qquad\Delta E_{i \rightarrow j} <0
\end{equation}
while for phonon absorption
\begin{equation}
f(\Delta E_{i \rightarrow j}) = \frac{1}{e^{\Delta E_{i \rightarrow
      j}/T}-1}, \qquad \Delta E_{i \rightarrow j} >0.
\end{equation}
We have made no estimates for the numbers $t_0$ and $T_0$, so for
the plots showing a time scale, they should both be taken as unity.
The time for one electron jump, $\Delta t$, is the inverse of the total rate 
\begin{equation}
\Delta t = \frac{1}{\Gamma_{tot}},\qquad
\Gamma_{tot}=\sum_{i \neq j}\Gamma_{i\rightarrow j}
\end{equation}
This is the time during which, on average, one jump takes place.
To get the correct noise spectrum at very short time scales, $\Delta
t$ should have been drawn from a distribution with average $1/\Gamma_{tot}$,
this we have ignored.

We have not seen any other simulations stating that they use the full
expressions for the jump probability, most authors seem to use the low
temperature limit of the expressions to reduce computational effort.
Temperature in all equations is the phonon bath temperature.

\section{Results of relaxation}
\begin{figure}
\includegraphics{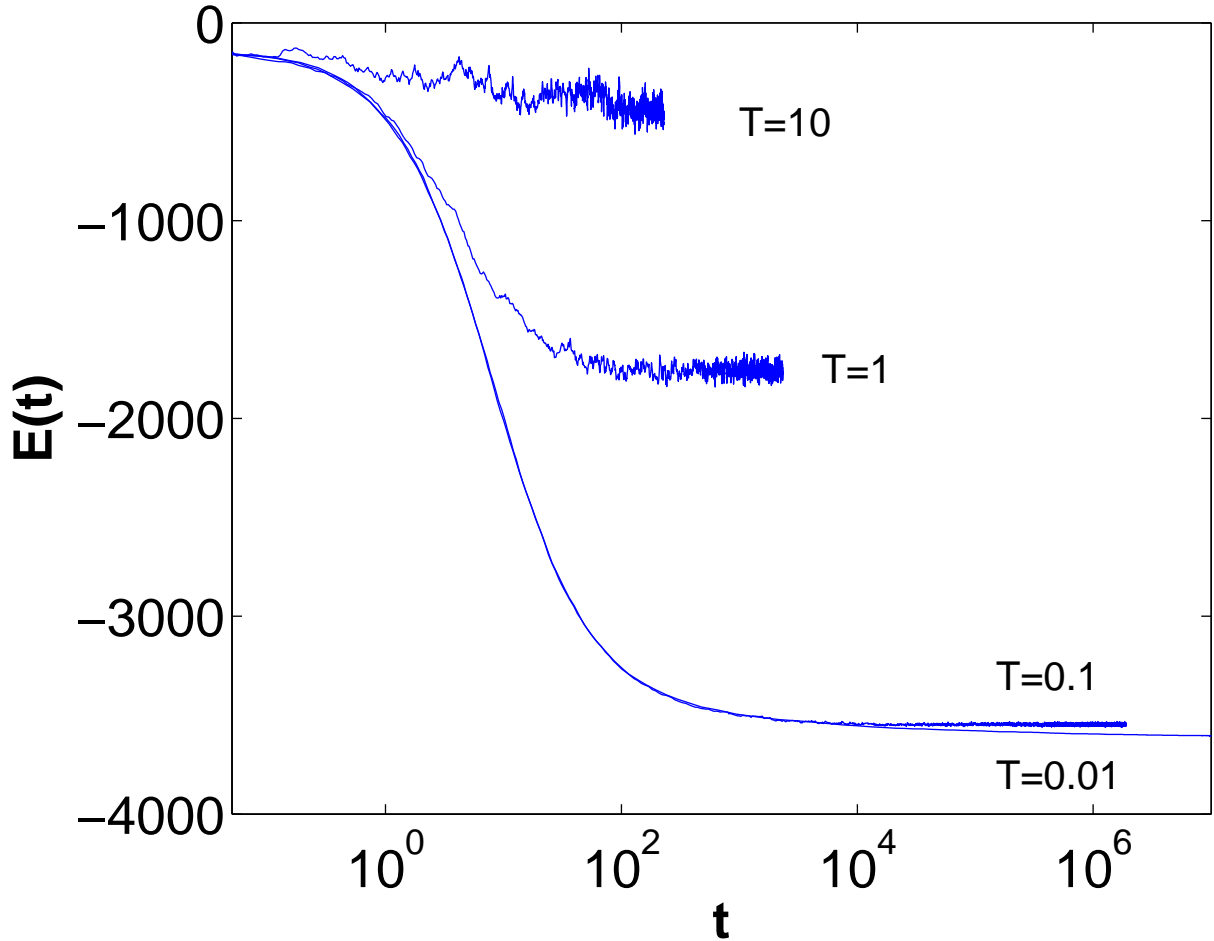}
\caption{\label{basicrel} Plot of the relaxation from $T=\infty$ at
  temperatures $T=\{10,1,0.1, 0.01\}$, the equilibrium energy
  decreases with $T$. We see that for $T=0.1$ we still get an
  equilibrium, while for $T=0.01$ or lower, equilibrium is not
  achieved at our time scale. Close-ups of the tails of the two lowest
  graphs are shown in Fig. \ref{zoomrel}}
\end{figure}

In order to study the temperature dependence of the relaxation
process, all relaxations were started from the same random electron
configuration, representing an infinite temperature and with the same
realization of disorder. We then set the phonon bath to a given
temperature, simulating a rapid quench.  We follow the relaxation 
process by plotting the total system energy as function of time, as
done by Ortu\~no et al \cite{Ortuno} for the model with positional disorder. 
 As shown in figure
\ref{basicrel}, the processes initially looks the same at all temperatures.
This is because the system is still at some high energy and does not ``feel'' 
the 
differences in phonon temperature which are small compared to the initial
infinite temperature of the system. In this regime, essentially all jumps
reduce the total energy of the system.

\begin{figure}
\begin{tabular}{ccc}
\includegraphics[width=5cm,height=4cm]{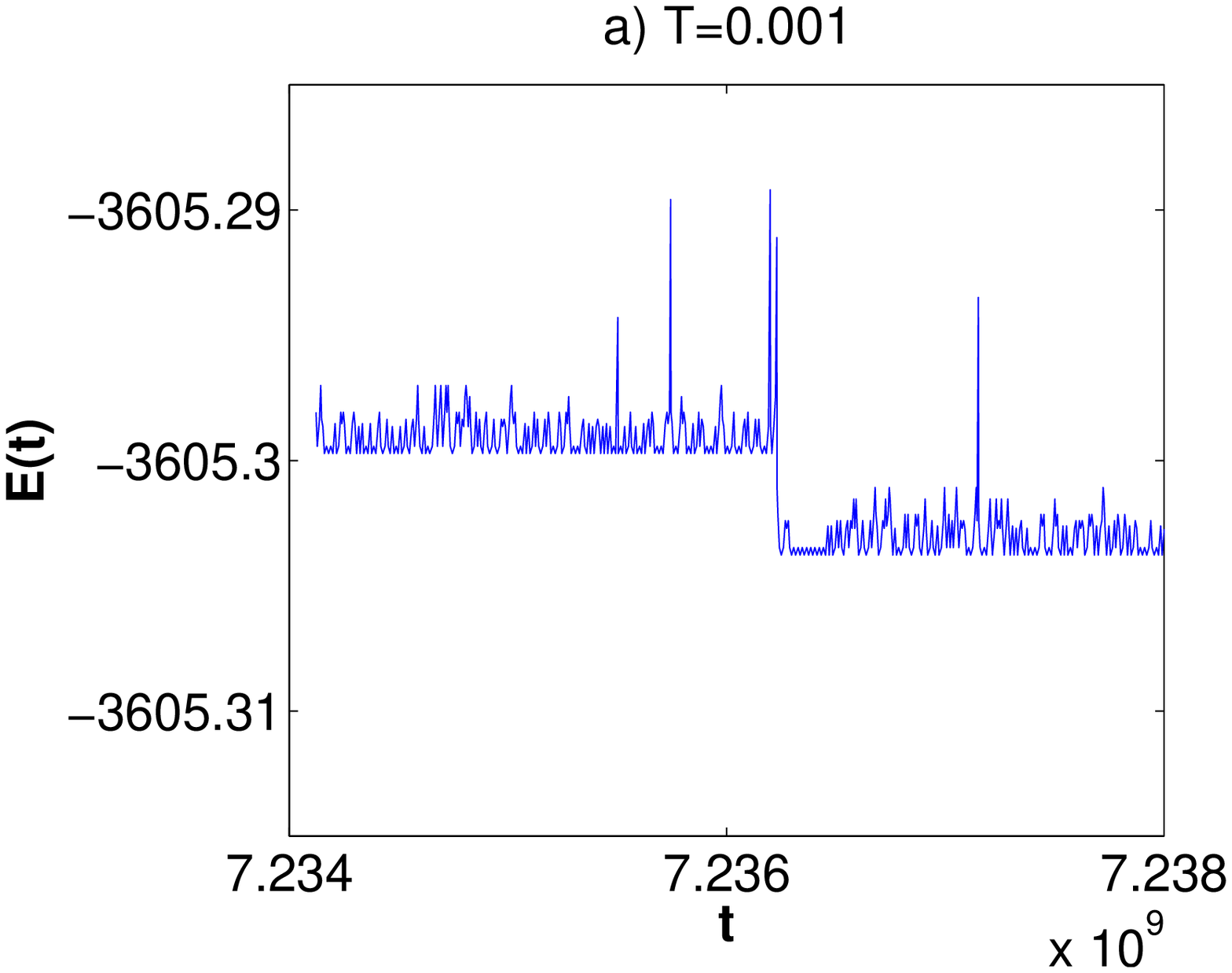}& 
\includegraphics[width=5cm,height=4cm]{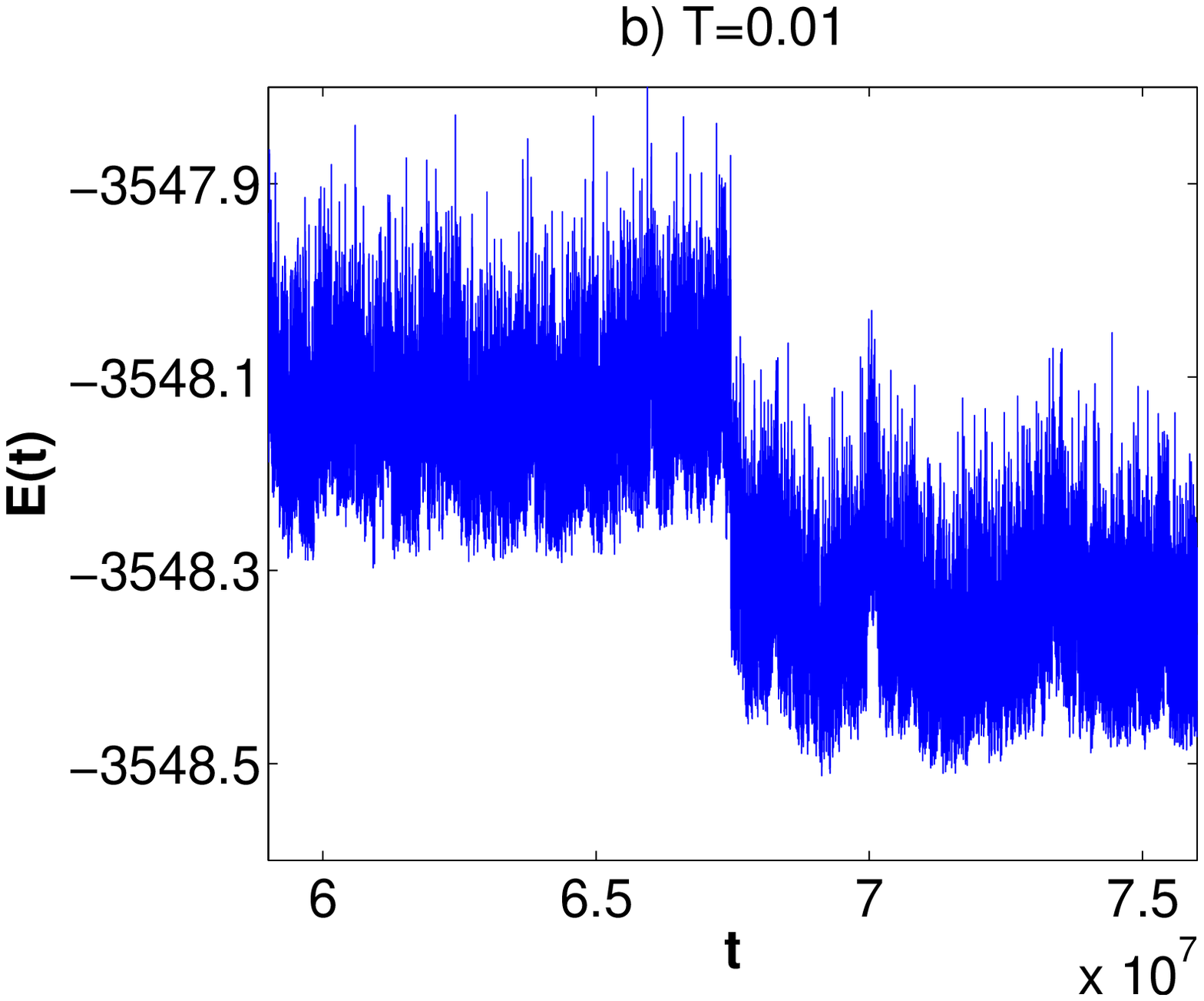} &
\includegraphics[width=5cm,height=4cm]{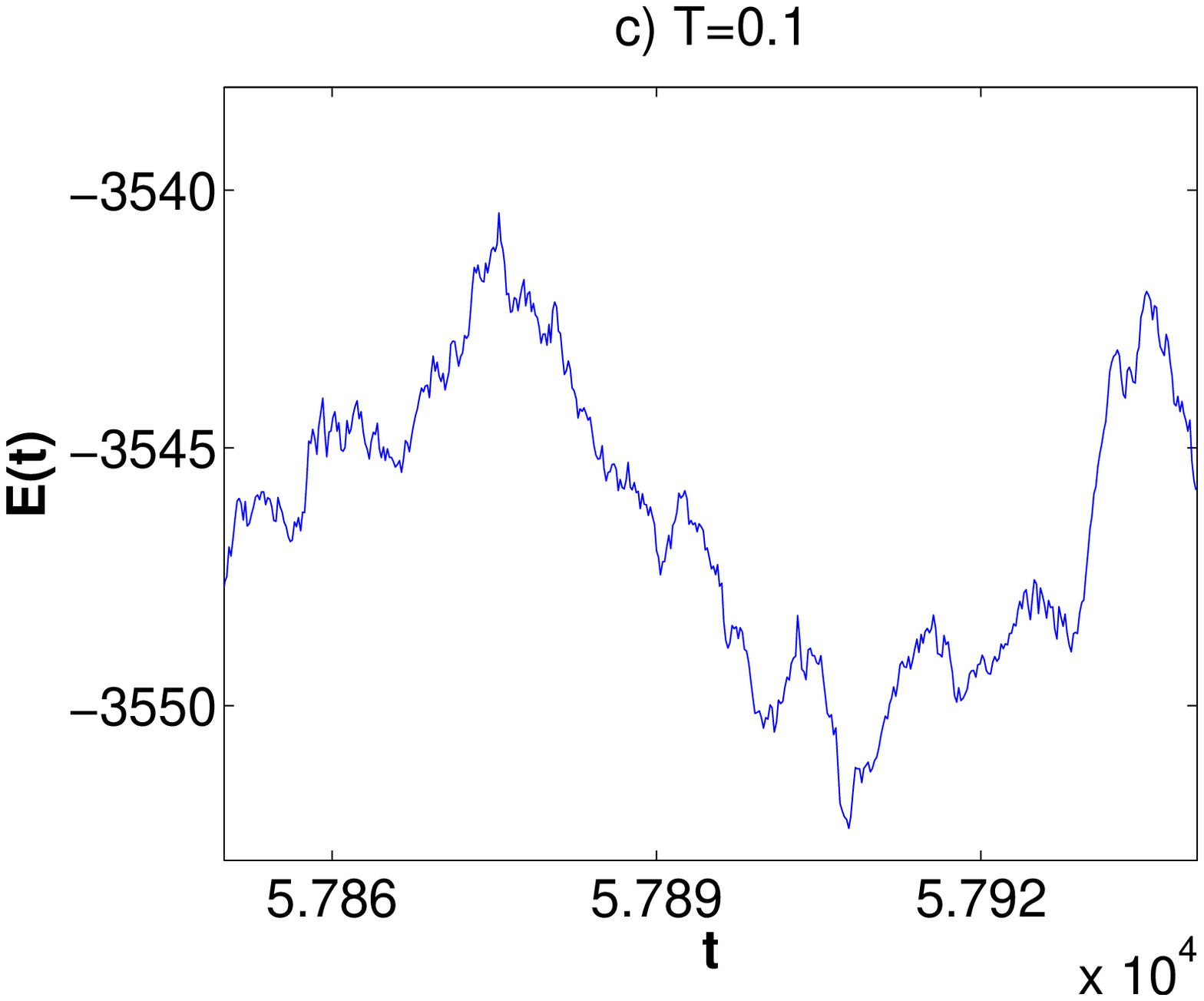}
\end{tabular}
\caption{\label{zoomrel} Close-up of the energy development at
  $T=\{0.001, 0.1\}$, 500 jumps each, and $T=0.01$, 250000 jumps.
  $T=0.001$ shows a clear local minimum performing steps, $T=0.01$ has
  no sharp lower limit, but shows both steps and a tendency of
  differences in going up and down, while for $T=0.1$ there seems to
  be good symmetry around the mean and no steps can be identified. The
  width of the distribution is many times the energy of single jumps,
  which are of the order $T$}
\end{figure}

Then, starting with the highest temperatures, one by one the systems
seem to reach an equilibrium behavior, with the energy fluctuating
around some average value. For all temperatures we observe that the
relaxation is slower than exponential.
 For the lowest temperatures ($T\lesssim0.02$)
we are not able to 
reach the equilibrium state, and the system continues to relax to lower
energies for our entire simulation time. 
Zooming in on the different relaxation graphs, we can see significant
qualitative differences, as shown in Fig. \ref{zoomrel}. At high
temperatures where equilibrium is reached ($T\gtrsim0.02$), we see
something resembling a random walk in a confining potential (Fig.
\ref{zoomrel}c).  For $T=0.01$ (Fig. \ref{zoomrel}b) we see periods
with something like a random walk around a central value, similar to
the high temperature equilibrium case. At intervals this is
interrupted by a sudden marked decrease in the energy, and the process
continues with a lower mean energy.  The periods of apparent
equilibrium can be understood as some local equilibrium around a
metastable local minimum of the energy and the steps correspond to the
system crossing to a different local minimum.  We have also observed
steps that increase the energy, but they are usually smaller and less
frequent, so that the long time average still is decreasing for the
whole simulation period.  For the lowest temperatures (Fig.
\ref{zoomrel}a), we see the energy making small fluctuations from a
clearly defined lowest level, which again decreases in clear
steps. The steps are now much larger than the width of the
distribution within one step. At these temperatures, the system
actually finds the local minimal state and spends a considerable
fraction of the time in this state. We believe that this is an effect
of the finite size of our system, and that if we had increased the
system size, the number of states accessible close to the local
minimum would also increase. The system would then not so easily find
its way to the local minimum, and we would get a picture similar to
Fig. \ref{zoomrel}b even at the lowest temperatures.  The nature and
statistics of the steps are discussed in greater detail in appendix
\ref{stepsappendix}.

\begin{figure}
\includegraphics{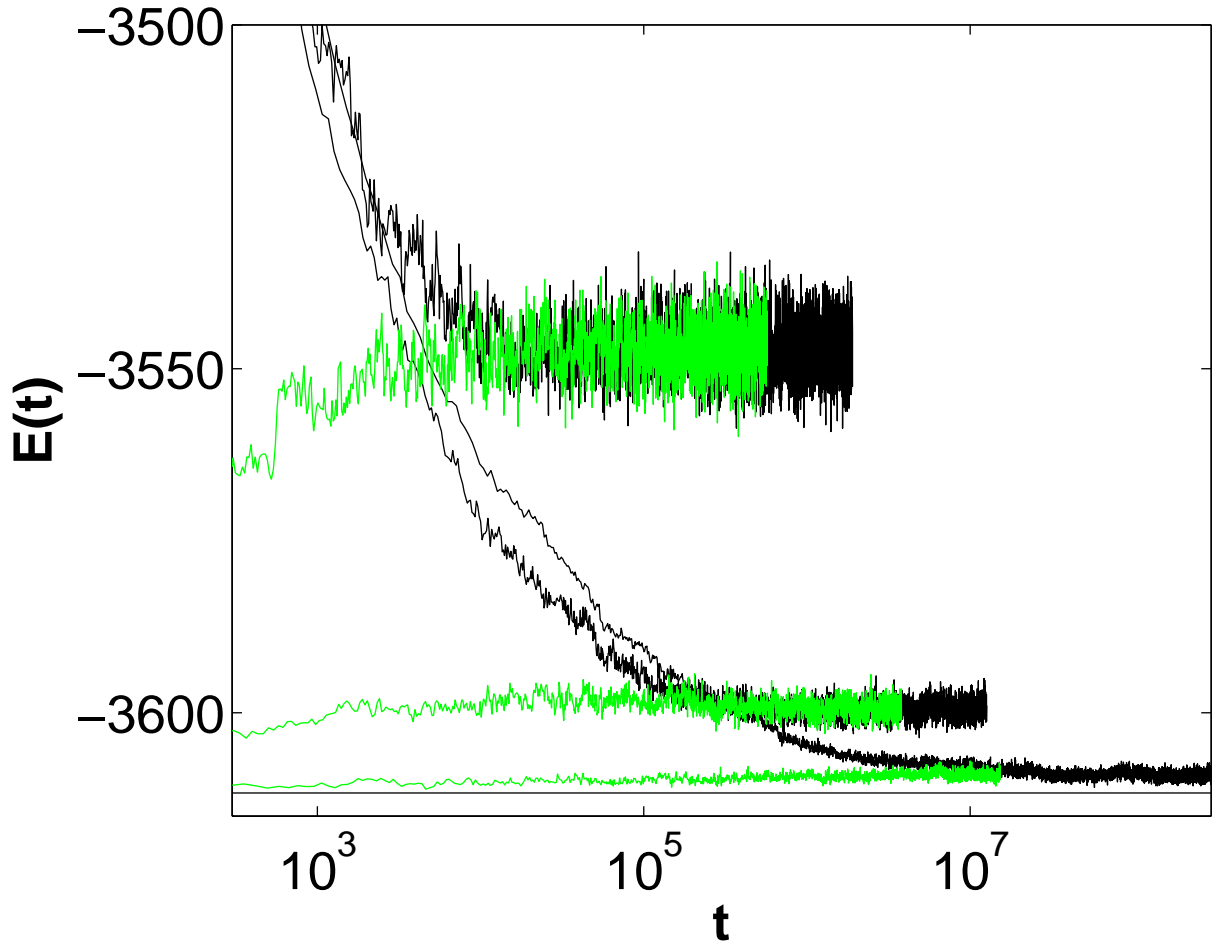}
\caption{\label{eqrel} Plot of the relaxation from high and low energy
  starting points for T=\{0.03, 0.05, 0.1\}. The straight line at the
  bottom shows the starting energy of the rising graph at $T=0.03$. For 
T=0.03 the graphs starting at high and low energy do not 
overlap fully because we have not continued the low-starting simulation 
for a sufficient number of steps. One can check that the average 
at the final part of both graphs are very close to the same.}
\end{figure}

One can ask whether the equilibrium that we seem to observe at temperatures
above $T=0.02$ really is a true equilibrium, or the system is still  
relaxing in energy, but so slowly that we are not able to see this in 
our simulations. 
In order to confirm that we have attained true equilibrium, we can
initiate the system at a low energy. We have used two different ways
of initiating the system. One is to relax the system at a lower
temperature, and then increase the temperature. The other is to relax
the system at zero temperature (using the algorithm described in
ref. \onlinecite{Glatz}, a less CPU-consuming procedure) a great number of
times, and picking the configuration with the lowest energy as the
starting energy for the Monte Carlo algorithm. Both of these were
tested, giving the same result. As shown in Fig. \ref{eqrel}, we see
that even though we initiate the system at a low energy, it relaxes
right up to the equilibrium energy range. We consider this as proof
that true equilibrium was reached at temperatures $T\ge0.03$.
$T=0.03$ is the lowest temperature for which we have been able to
achieve such a confirmation of equilibrium, simply because we have not
found any state with $E$ lower than at the end of the $T=0.02$ simulation. Thus
we have no low-energy starting point - at even lower
temperature, the relaxation is so slow that we have not reached energies
lower than those of the $T=0.02$ simulation. 

One could still worry that the system does not reach a full equilibrium, 
but instead the configuration space breaks into several ergodic 
components. In order to test this we started the system in several 
different initial configurations, both at high and low energy, thereby 
possibly ending up in different ergodic components if they exist. In all 
cases we observed that the system reached the same final average energy, 
thus indicating that we have achieved full equilibrium. 

\begin{figure}
\includegraphics{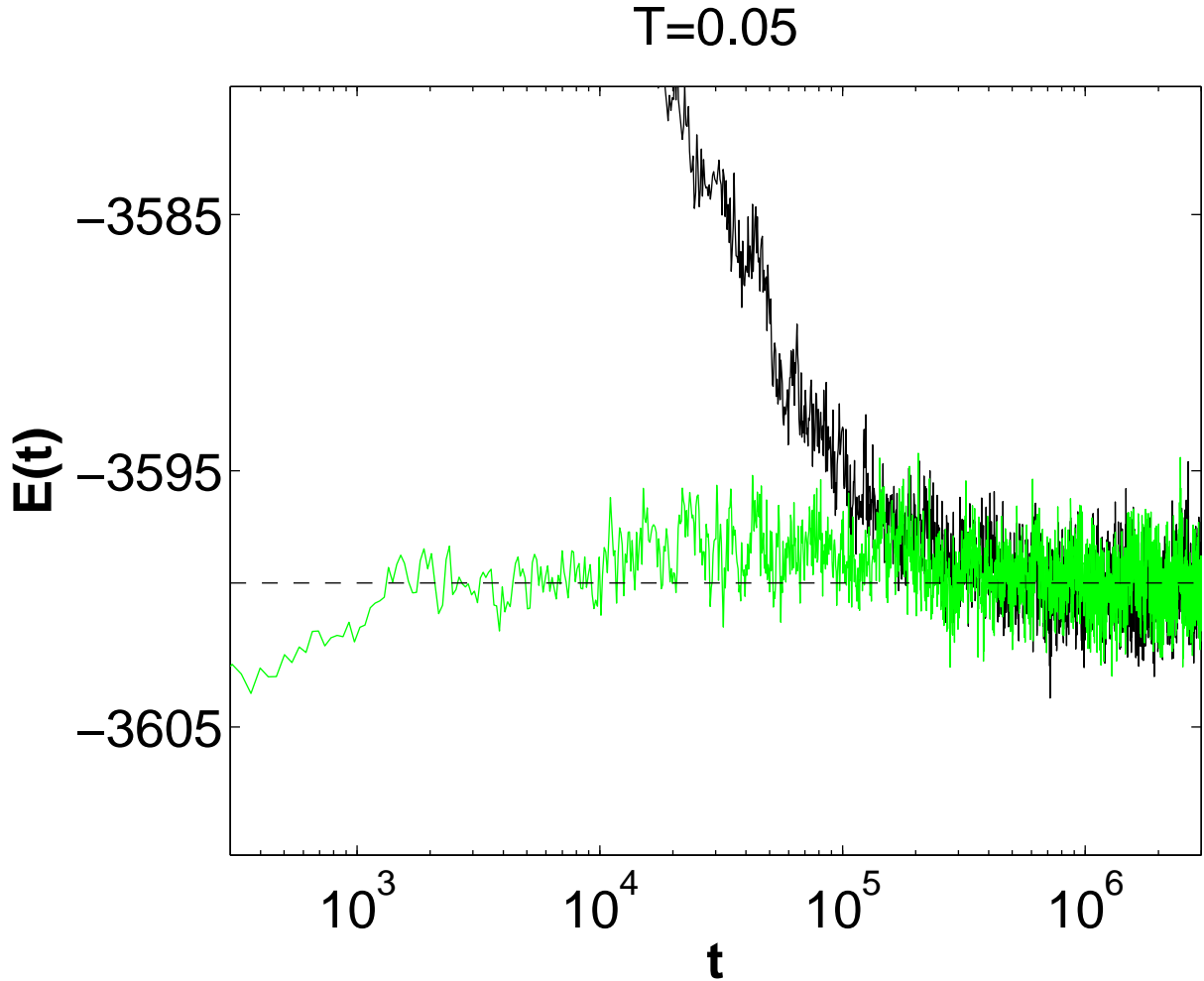}
\caption{\label{medtemprel} Plot demonstrating the difficulties in
  defining a clear time for whether and when equilibrium is reached.
  The dotted line represents the mean of the last half of the data
  points for the falling graph.}
\end{figure}

If we believe that there exists some glass transition temperature
$T_g$, below which the system will show typical glassy behavior, we
could expect that the time needed to reach equilibrium would diverge
at this temperature. Just looking at the energy relaxation graphs it
is difficult to decide when equilibrium is reached. A better idea is
to combine the graphs starting from high and low energy and use the
point where they start overlapping as a measure of the time needed to
reach equilibrium. As seen in Fig. \ref{eqrel} this can at least be
used to get an order of magnitude estimate.  However, the method can
never be very accurate for several reasons.  First, because of the
combination of low and high frequency thermal noise, the time when the
two graphs starts overlapping is not well defined. This is illustrated
in Fig. \ref{medtemprel}. Second, the result will clearly depend on
the initial states chosen for the high and low starting energies
respectively.

As an alternative, we can study the energy correlations of the system 
after it has reached equilibrium. We believe that the same timescales 
should be present both in the correlation function and in the final 
part of the relaxation process. 

\section{Analysis of system at equilibrium}

\subsection{Energy correlations}
\begin{figure}
\includegraphics{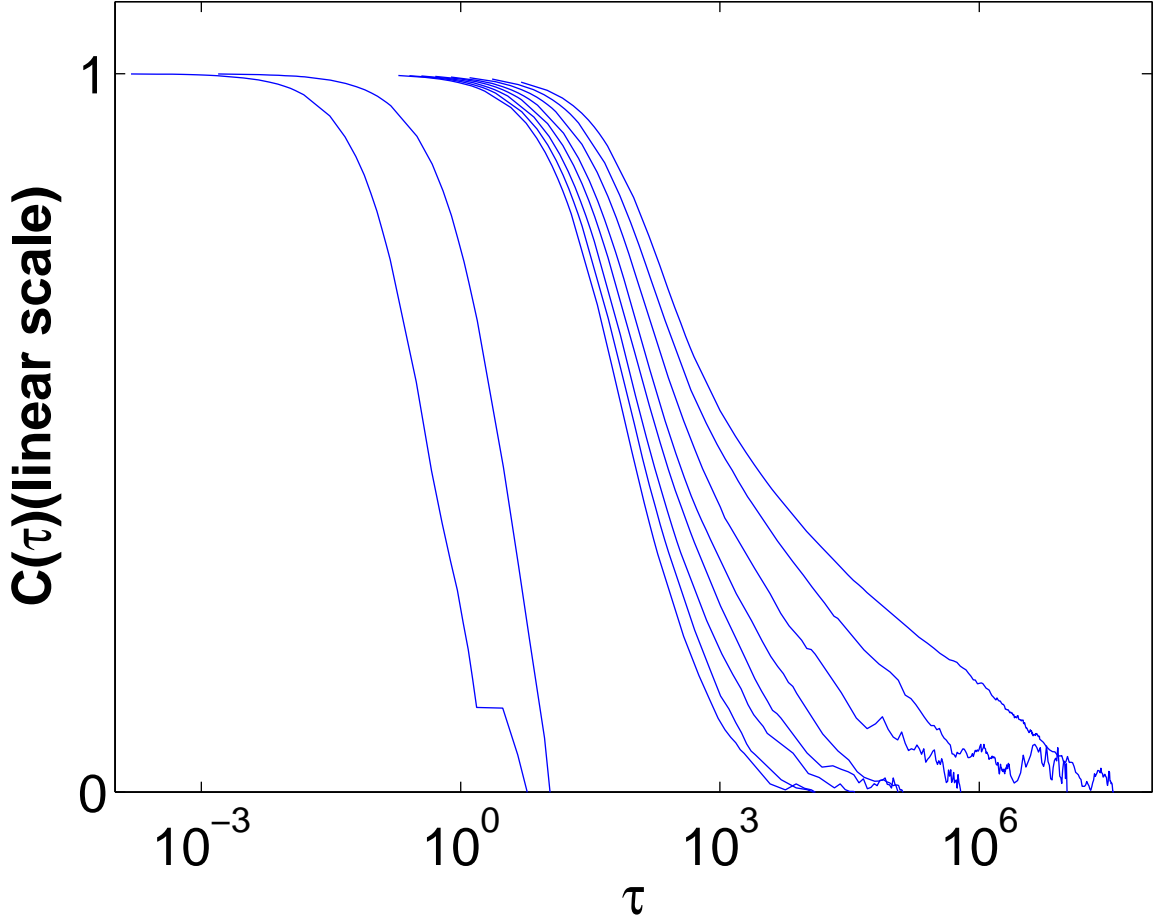}
\caption{\label{Cbasic} Plot of $C(\tau)$ at 
$T=\{0.03, 0.04, 0.05, 0.06, 0.07, 0.08, 0.09, 0.1, 1, 10\}$}
\end{figure}

The two-time energy correlation function is defined as
\begin{equation}
\label{corr2t}
C(t,t+\tau)=\frac{1}{\sigma^2} \left<(E(t)-\bar{E})(E(t+\tau)-\bar{E})\right>
\end{equation}
where $\bar{E}$ is the average energy and $\sigma$ is the standard
deviation of the distribution of energies. The average should be over all
realizations of the equilibrium dynamics.  When studying the system at
equilibrium, we use the fact that the system is stationary, and the
correlation function will depend only on the time difference $\tau$.
We can then use one simulated time evolution and average over starting
times $t$ instead of over different realizations of the system, and
write
\begin{equation}
\label{corrtau}
C(\tau)=\frac{1}{\sigma^2} \left<(E(t)-\bar{E})(E(t+\tau)-\bar{E})\right>_t.
\end{equation}
The average should include a large number of uncorrelated times, and this
is only true if the period we average over is
longer than the time scale of the correlations. Therefore the value of the
correlation function for time differences $\tau>10^6$ should not be
trusted for our time series of $10^7$ steps. Also, for long $\tau$ the
correlation function shows a lot of noise, but interesting effects can
be demonstrated at significantly shorter times than this.
$C(\tau)$ at a number of temperatures is plotted in figure \ref{Cbasic}.

Kolton et al. \cite{Kolton} made a similar analysis of the occupation 
correlation function for the three dimensional random site model. 
They proposed that one 
should plot the curves as function of the scaled variable $c_T t$ 
where $c_T$ is some temperature dependent relaxation rate. 
Based on a previous study\cite{Grempel} they assumed an activated 
law $c_T = e^{-T_0/T}$. We wanted to avoid this assumption, and instead
extract the relevant rate from our numerical data using the following 
reasoning. 

\begin{figure}
\begin{tabular}{cc}
a) & b) \\
\includegraphics[width=7cm]{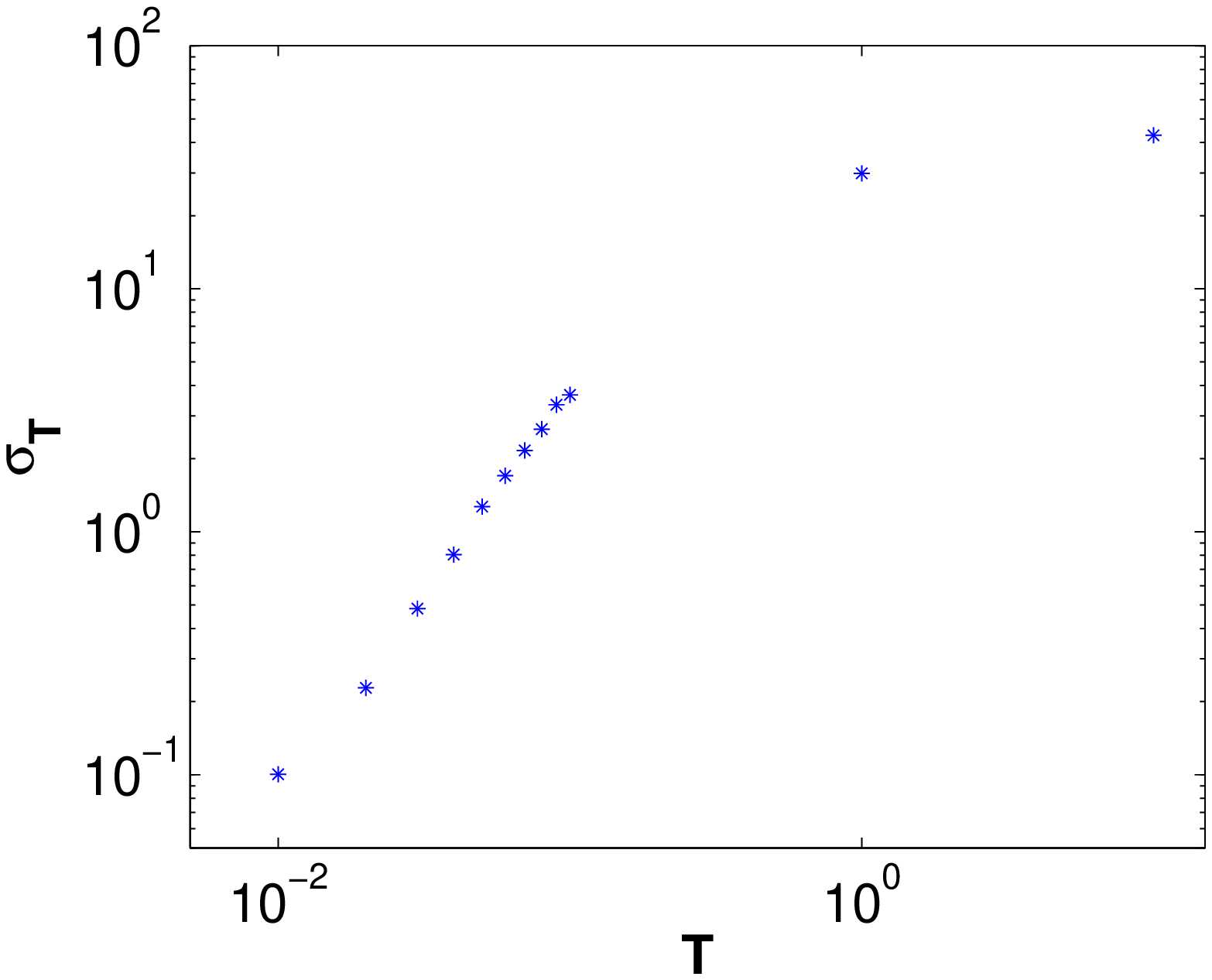} &
\includegraphics[width=7cm]{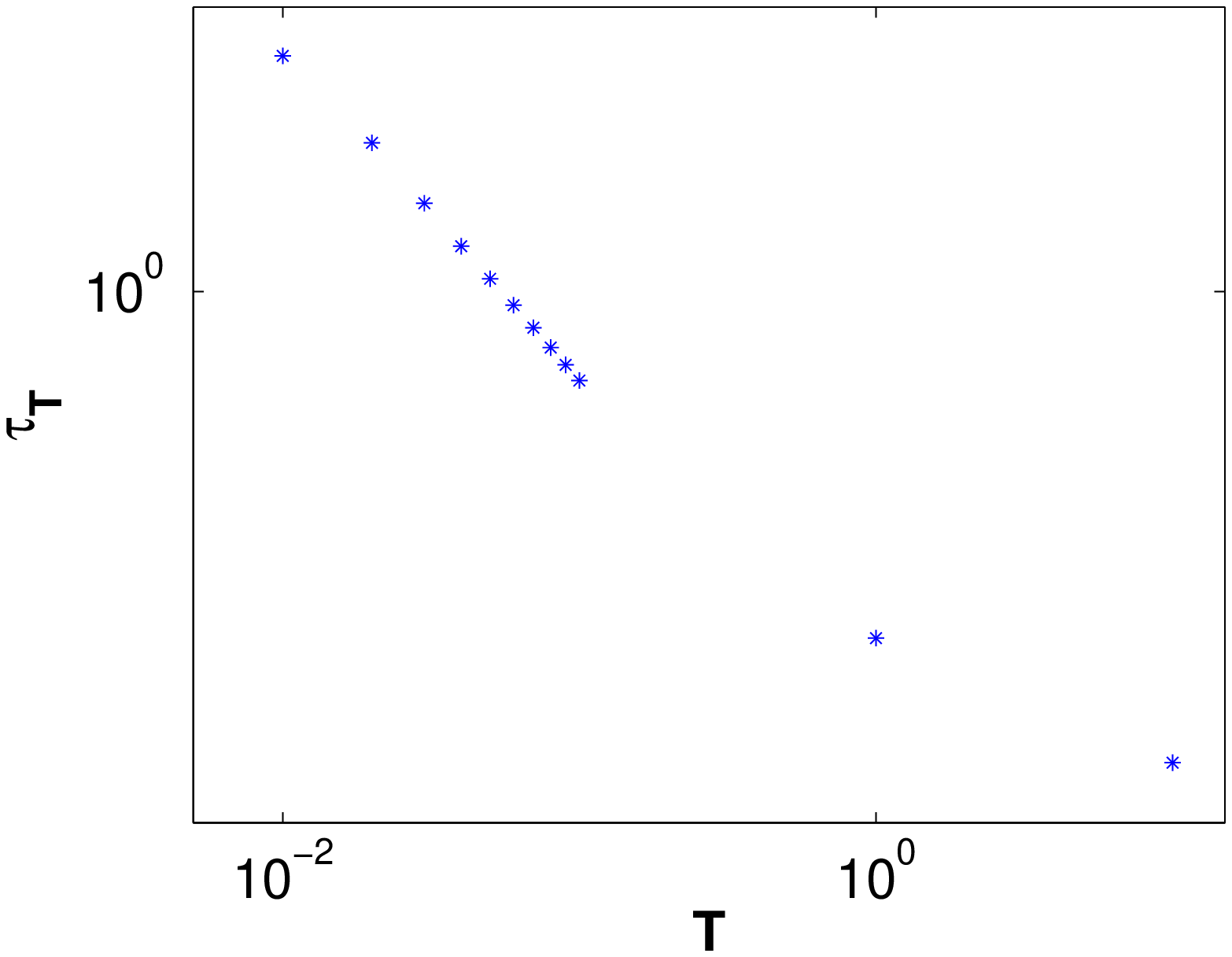} \\
c) & d) \\
\includegraphics[width=7cm]{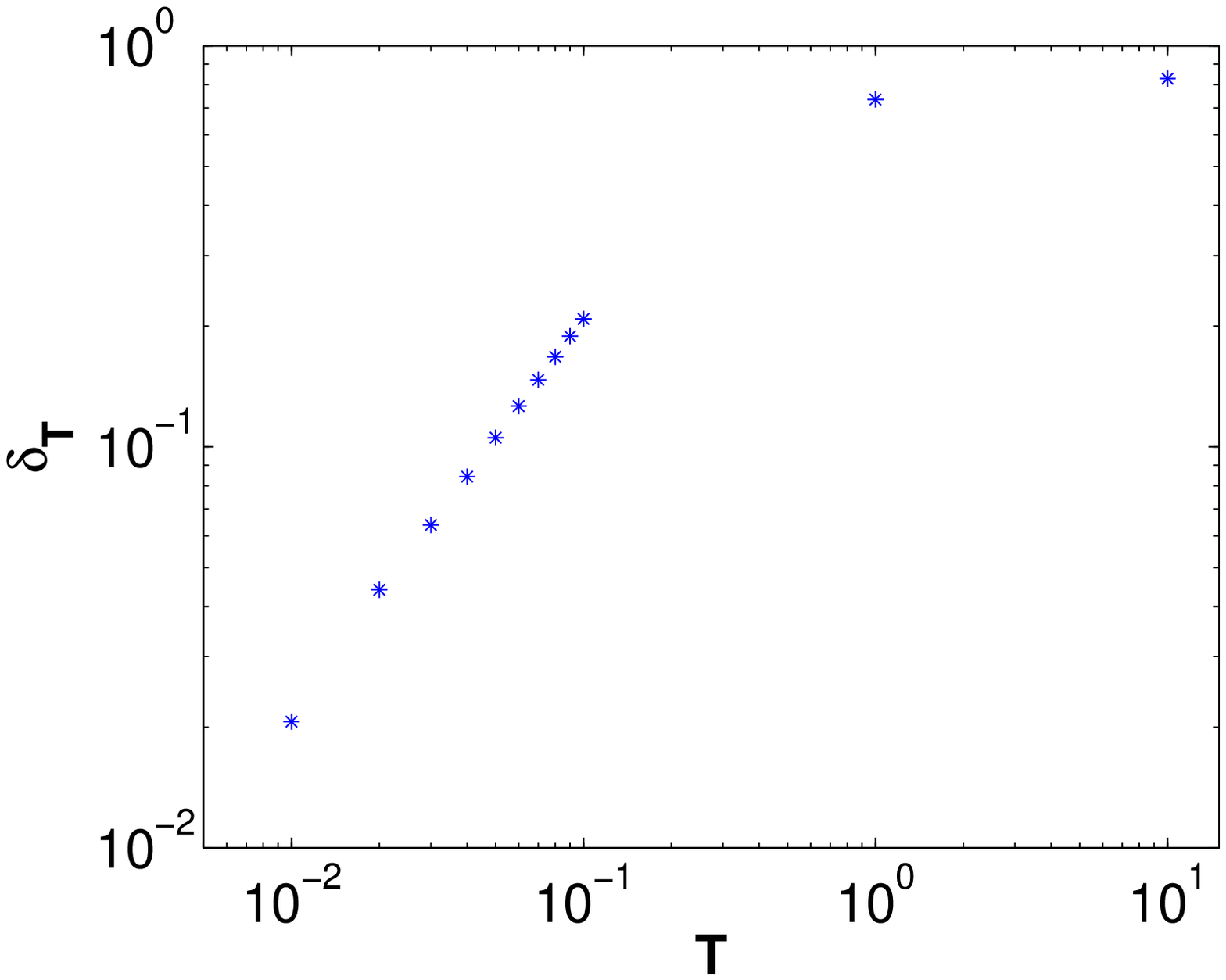} &
\includegraphics[width=7cm]{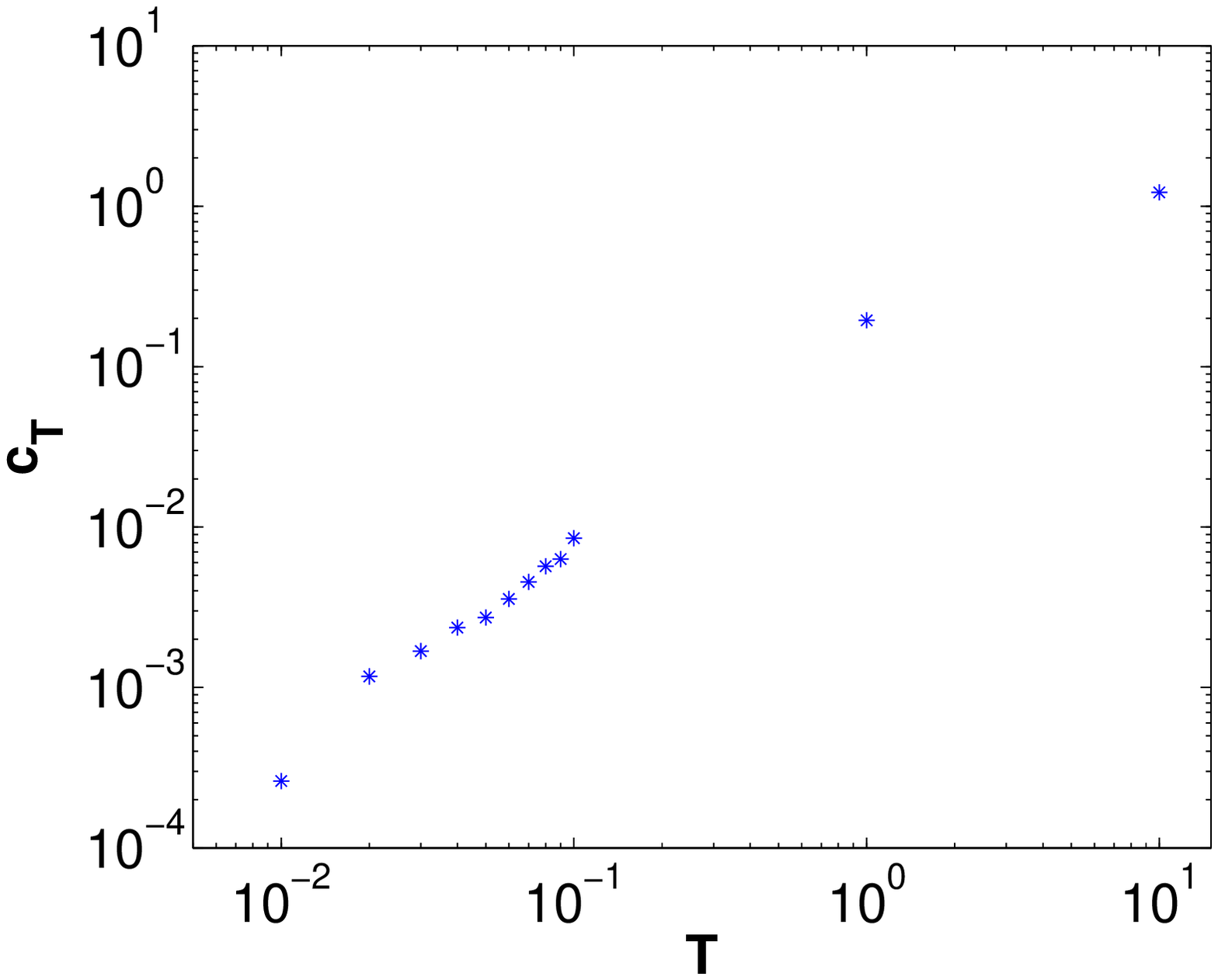} 
\end{tabular}
\caption{\label{params} a) $\sigma_T$, width of the energy distribution b) $\tau_T$, average time per step c) $\delta_T$, mean of $|\Delta E_{i \rightarrow j}|$ d) $c_T=\frac{\delta_T^2}{2 \sigma_T^2 \tau_T}$}
\end{figure}

We try to model the evolution of the system as a random walk in
energy.  The probabilities of increasing or decreasing the energy at a
particular step of the random walk is given by a combination of the
transition rates given in Eq. (\ref{tunnelingrate}), and the density
of states. It can be shown that this is equivalent to the problem of a
random walk in a harmonic potential. The details of the derivation are
given in appendix \ref{randomwalker}, the main results are that we
expect the correlation function to decay exponentially, $C(\tau) =
e^{-c_T\tau}$, with
\begin{equation}
\label{cT}
c_T=\frac{\delta_{T}^2}{2\sigma_T^2 \tau_T},
\end{equation}
where $\delta_{T}$ is the mean of the absolute value of the energy
change per electron jump at temperature $T$, $\sigma_T$ is the
standard deviation of the distribution of the system energy at the
same temperature, while $\tau_T$ is the average time per step. All
these parameters are accessible from our simulations.  Plots of the
relevant $\sigma_T$, $\delta_{T}$, $\tau_T$ and $c_T$ are given in
figure \ref{params}. 

For the $\delta_{T}$, we see that it simply rises as $\delta_{T}=2T$,
until the temperature becomes comparable with the spread of the single
particle energies. $\sigma_T$, $\tau_T$ and $c_T$ all give fairly
straight lines in the log-log plots, but we do not have sufficient
data to conclude what functional dependency they follow. However, the
previous suggestion of an activated law for $c_T$ does not seem to
fit our data.

The interpretation of $c_T$ is also clear. Since the variance of a 
random walk grows linearly in time, $1/c_T$ is the time needed for the 
random walk to spread over a range equal to the equilibrium 
standard deviation, $\sigma_T$. In other words, it is the time at which 
the random walk starts to feel the effect of the constraining potential. 
While $c_T$ includes the size-dependent quantities
$\sigma_T$ and $\tau_T$, it can be assumed that $c_T$ itself is size
independent, as $\tau_T \propto N^{-1}$ while $\sigma \propto
N^{1/2}$. 

\begin{figure}
\begin{tabular}{cc}
a) & b)\\
\includegraphics[width=7cm]{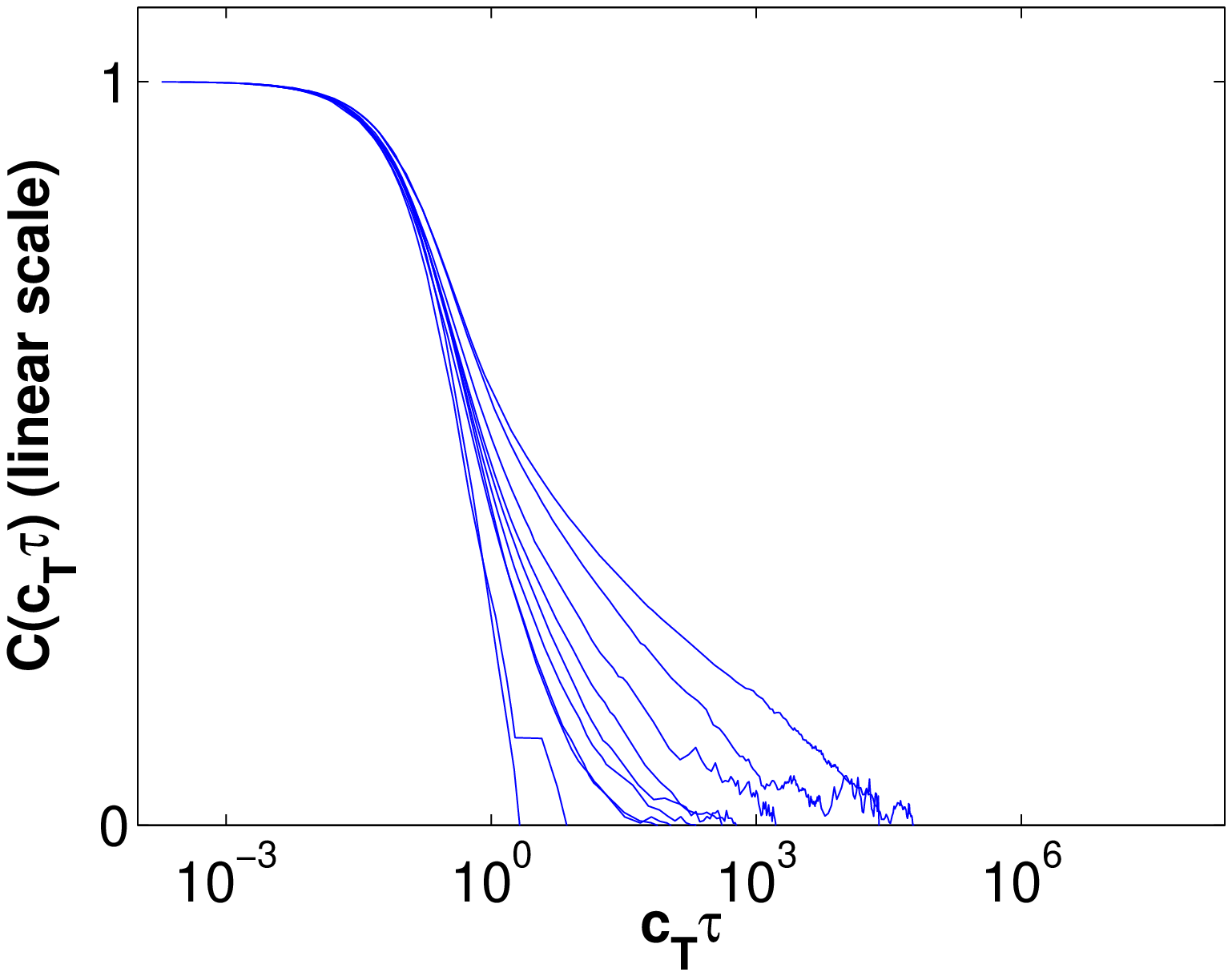}&
\includegraphics[width=7cm]{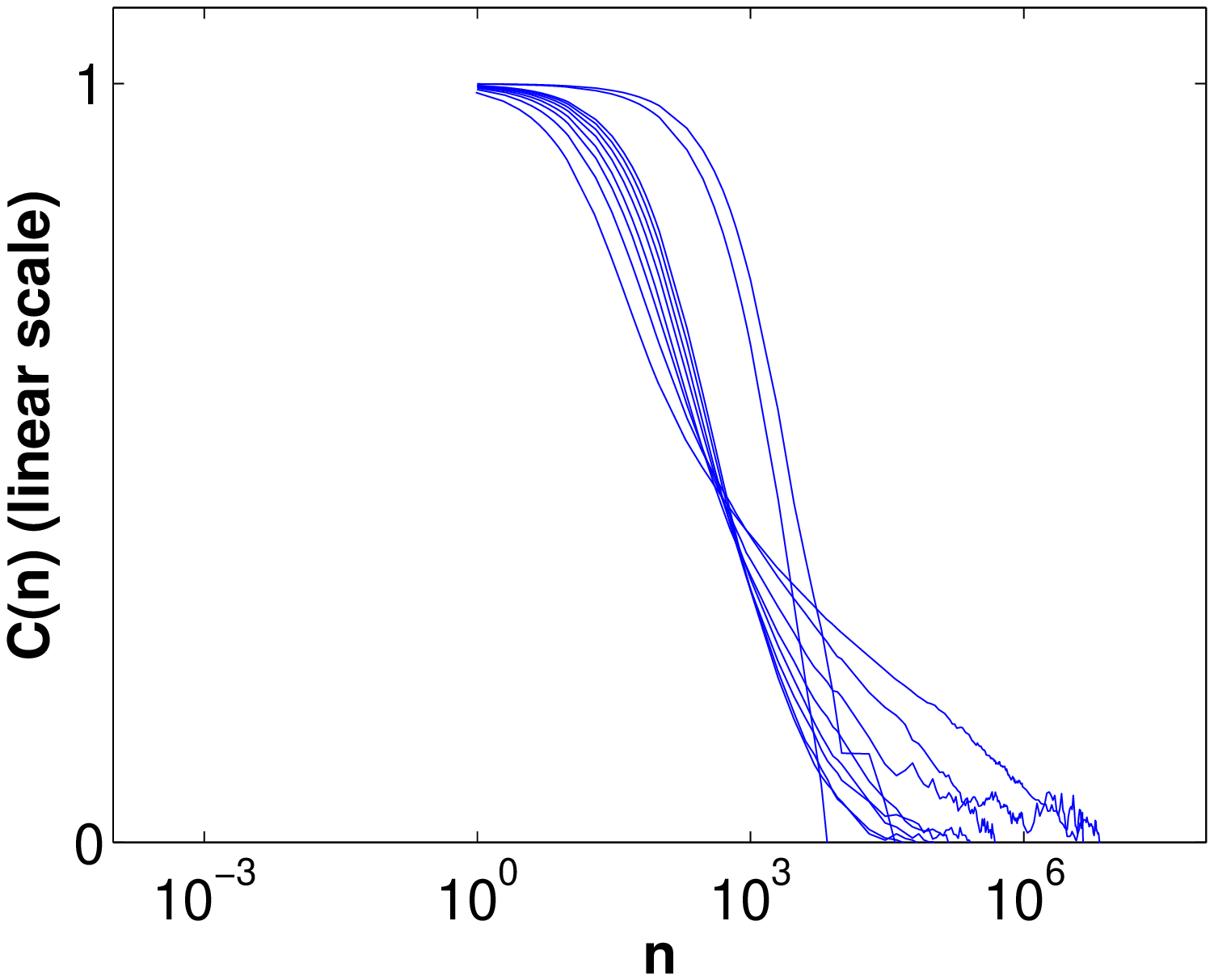}
\end{tabular}
  \caption{\label{Ccoll} Attempts at collapsing the correlation
    functions. T=\{0.03,0.04,0.05,0.06,0.07,0.08,0.09,0.1,1,10\} 
    a) Scaling by $1/(2c_T)$.
    High temperatures correspond to steep slopes
    b) $C(n)$, where $n$ is the number of
    steps, regardless of time spent. $T=\{1,10\}$ lie far away from the others.
}
\end{figure}

Fig. \ref{Ccoll}a shows the scaled correlation functions $C(c_T\tau)$, 
and as can be seen this gives an excellent
collapse for the initial stage of the relaxation, but very poor for
longer $\tau$. 

Inspired by the apparent success of Kolton et al. \cite{Kolton} in
collapsing the curves, we made several attempts with other rescaling
factors.  The most successful was using $\tau_T$ as scaling factor,
simply plotting the correlation as function of number of steps
performed, as shown in Fig. \ref{Ccoll}b. We see that we get a picture
resembling the collapse demonstrated by Kolton et al. for the random
position model. But the deviations from a collapse are systematic, and
by closer inspection of the inset of Kolton's Fig. 2. we get the
suspicion that their choice of zoom hides the fact that they have no
true collapse either, as the difference between the curves in their
scaled plot is an order of magnitude in time near both $C=1$ and $C=0$.

\begin{figure}
\begin{tabular}{cc}
a) & b)\\
\includegraphics[width=7cm]{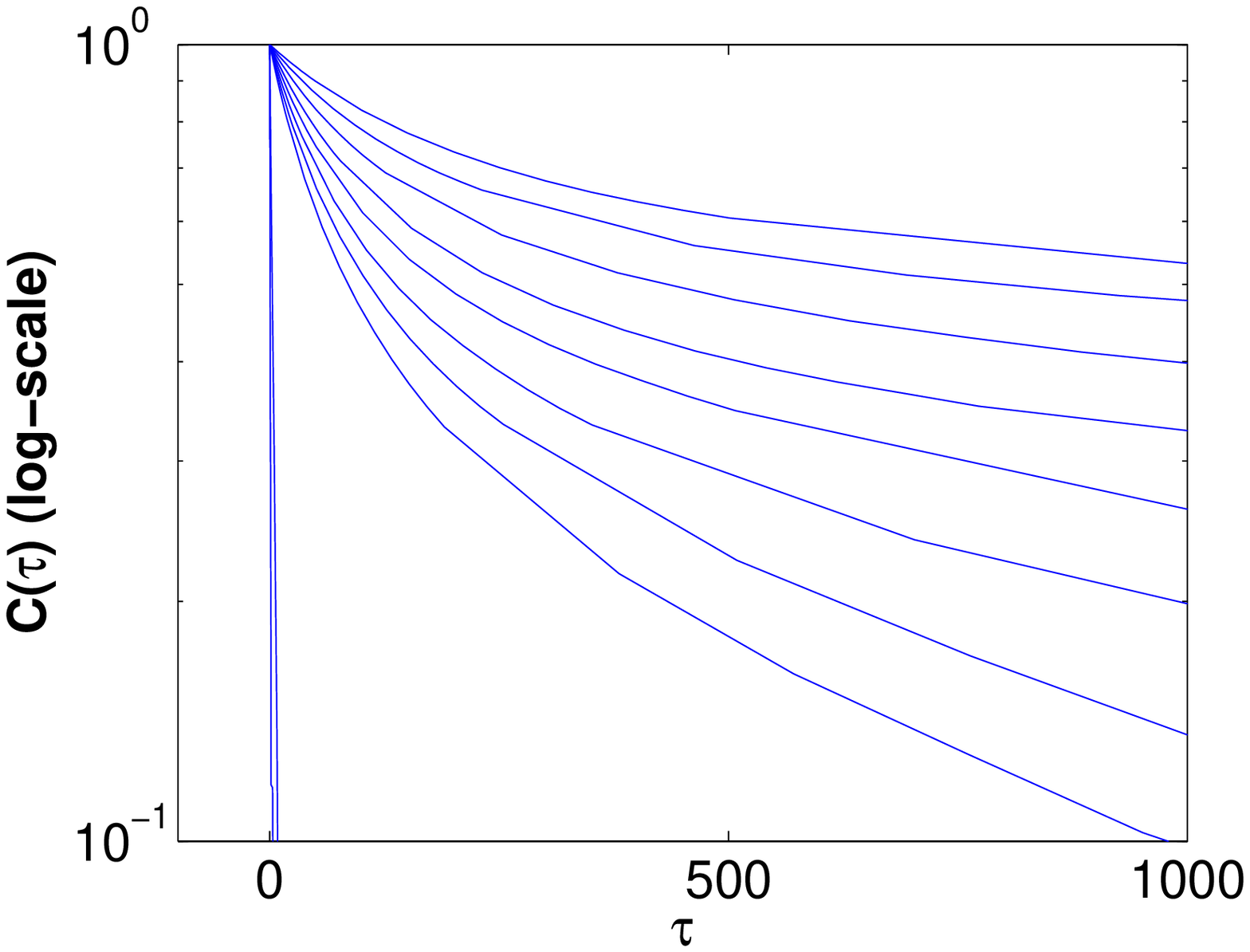} &
\includegraphics[width=7cm]{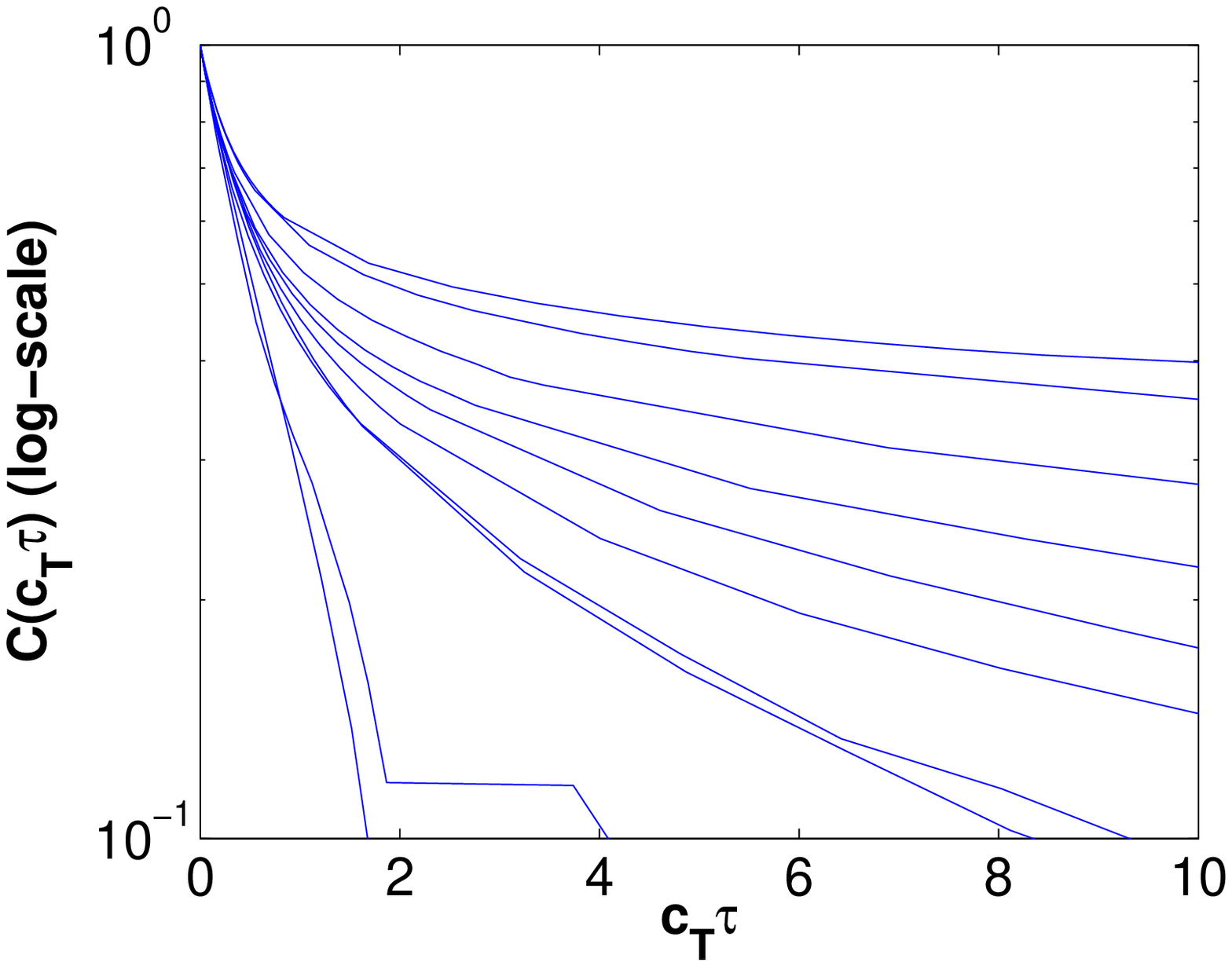}
\end{tabular}
\caption{\label{logC} a) $C(\tau)$ b)$C(c_T\tau)$, 
  $T=\{0.03-0.1,1,10\}$, we see signs of collapse at very short times, but only the highest temperatures give close to straight lines.} 
\end{figure}

If we instead use the assumption of the exponential decay and plot
$\ln(C(\tau))$ versus time, we expect to see straight lines. Figure
\ref{logC} shows that this does indeed come close to the truth for
$T=1,10$, but for the lower temperatures the lines are far from
straight. Only at short times one might think that there can be some
exponential decay.  This can to some extent be further justified by
plotting the scaled graphs $\ln(C(c_T \tau))$ (Fig. \ref{logC}b).  We
see that the high temperature (T=1,10) graphs are close to straight
lines whereas at lower temperature the graphs show a decay slowing
with time, as longer and longer time scales come into play.  At short
times they approach the straight lines defined by the high temperature
graphs, indicating short time exponential behavior, with the predicted $c_T$. 

\begin{figure}
\includegraphics{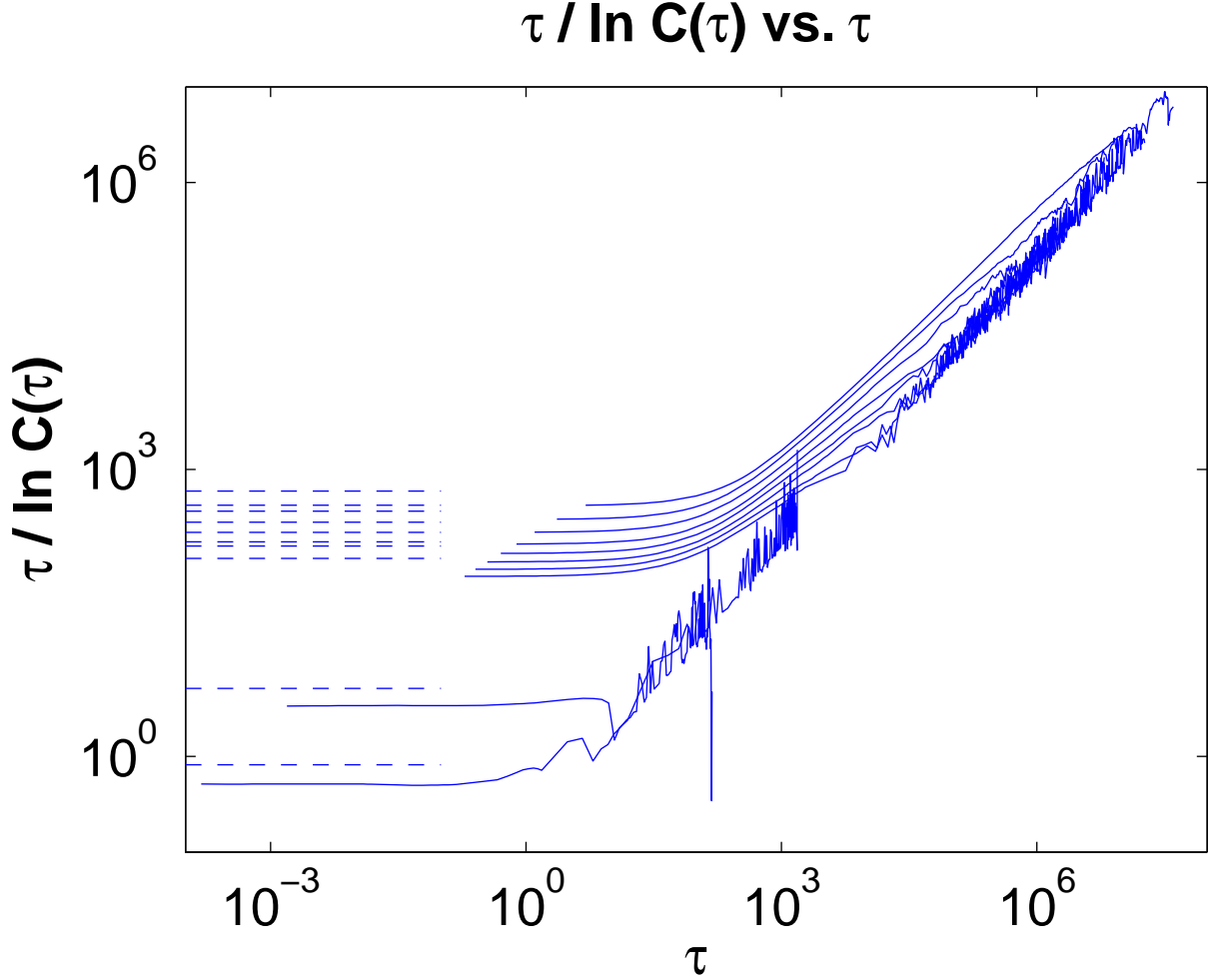}
\caption{\label{ttologC} T=\{0.03,0.04,0.05,0.06,0.07,0.08,0.09,0.1,1,10\} Horizontal lines indicate exponential behavior, slope of 1 indicates noise, while straight lines at other slopes indicate stretched exponential behavior. The dotted lines indicate $1/c_T$, as calculated using Eq. (\ref{cT}).}
\end{figure}

\begin{figure}
\begin{tabular}{cc}
a) & b)\\
\includegraphics[width=7cm]{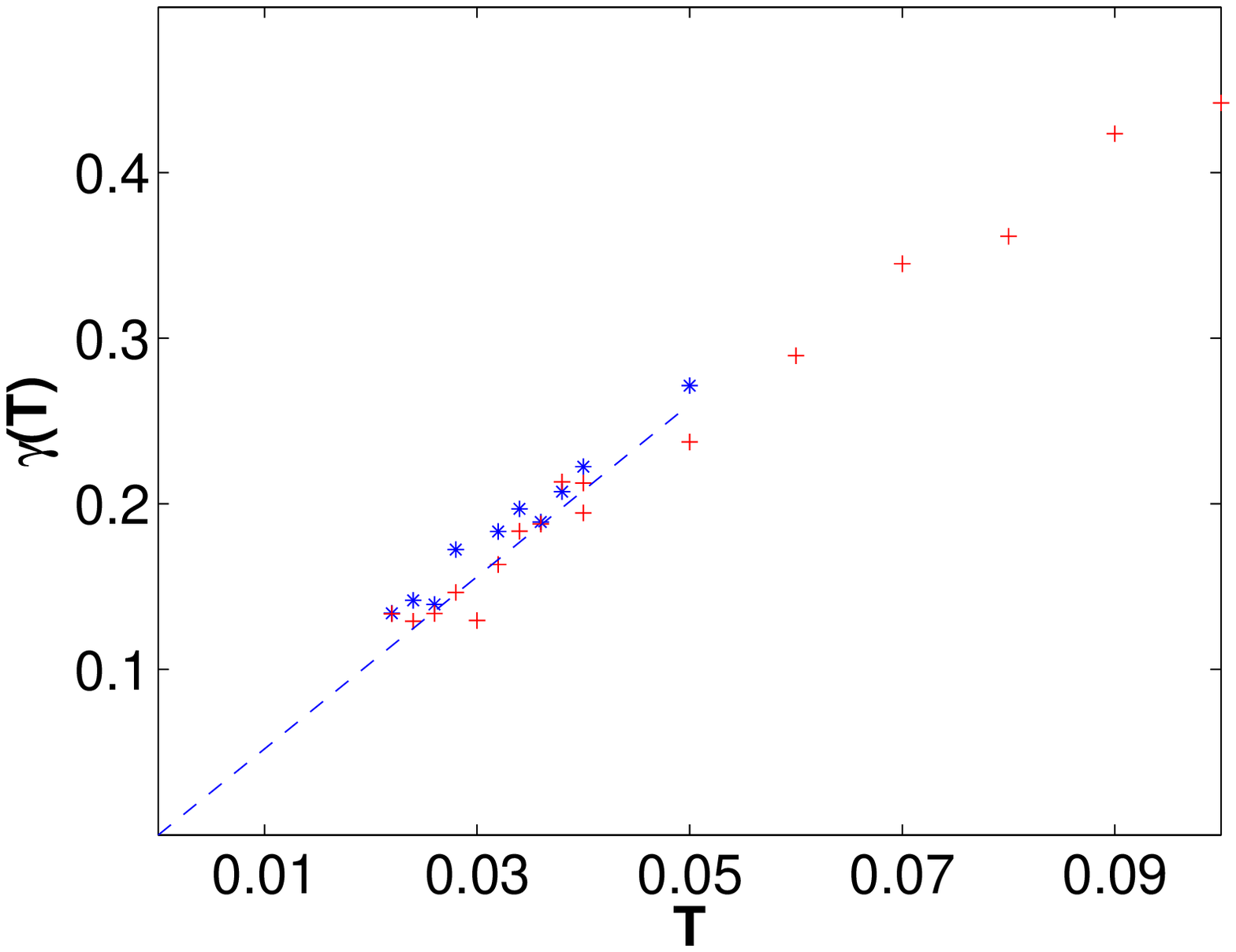}&
\includegraphics[width=7cm]{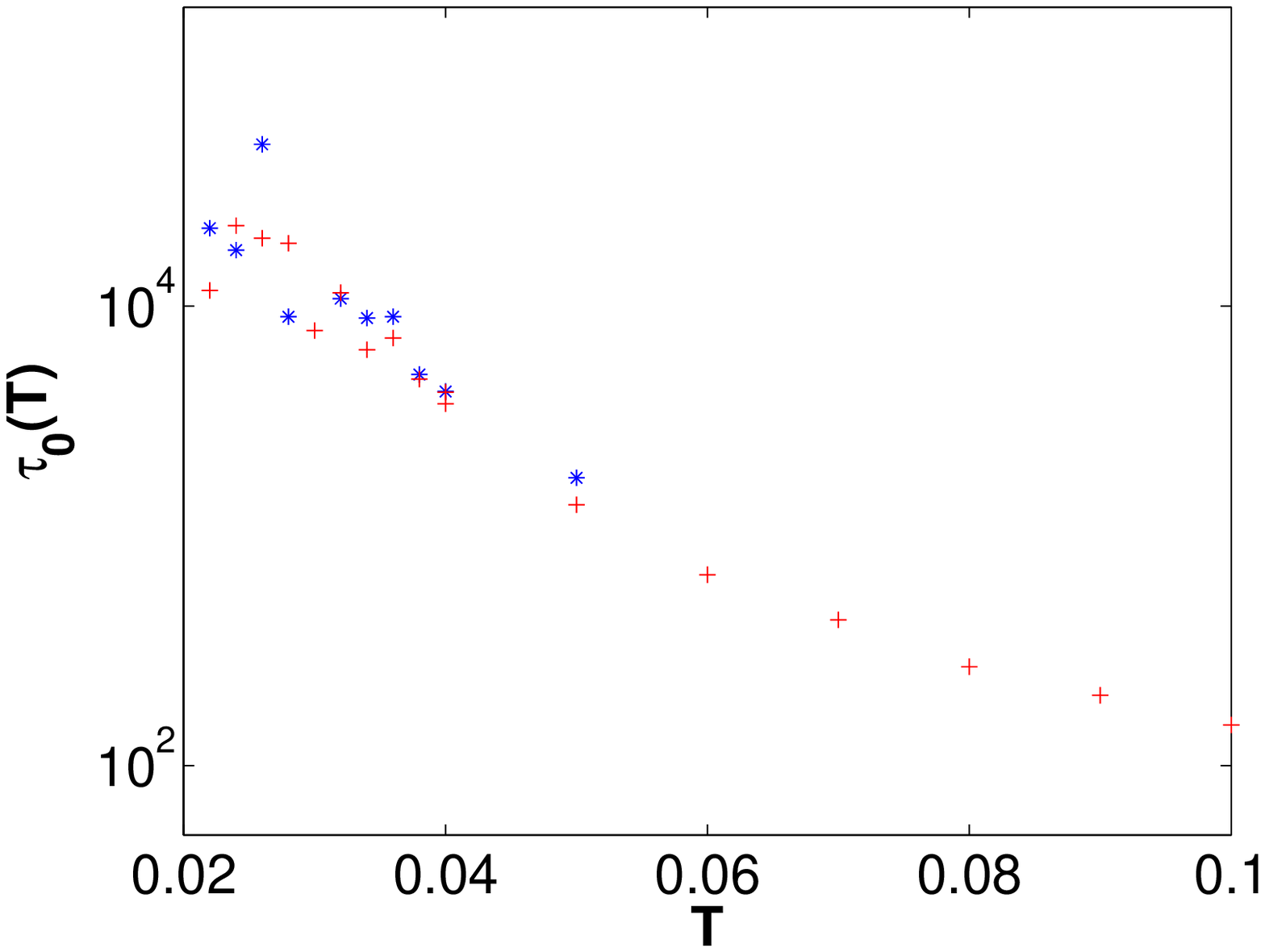}
\end{tabular}
\caption{\label{gammatau} a) $\gamma(T)$ b) $\tau_0(T)$ The dashed
line indicates a possible extrapolation towards $T=0$. The difference
between the symbols denotes use of different cut-off limits when
calculating jump rates, as discussed in appendix \ref{algorithm}. (*)
denotes a limiting value $\Gamma_{min}=10^{-10}\Gamma_{tot}$, (+)
denotes $\Gamma_{min}=10^{-7}\Gamma_{tot}$. For $\tau_0$ this makes no
significant difference, but for $\gamma(T)$ we see that the more
accurate simulation gives a slightly higher value of $\gamma(T)$. This
is the only graph where the choice of limiting rate is noticeable, confirming that our algorithm is sufficiently accurate.}
\end{figure}

In disordered systems, it has previously been observed\cite{Jund,Ogielski}
that correlation functions can be well fitted by a stretched exponential  
function  
$C(\tau)=A e^{-(\tau/\tau_0)^\gamma}$. If that is true, 
\begin{equation}
\ln{\frac{-\tau}{\ln(C(\tau))}}=\ln\frac{-\tau}{\ln A - (\tau/\tau_0)^\gamma}\approx (1-\gamma) \ln \tau + \ln\tau_0^\gamma,
\end{equation}
for $\tau \gg \ln(A)$, and plotting $\ln \tau$ vs
$\ln(-\tau/\ln C(\tau))$ we should get a straight line for large
$\tau$. For a pure exponential, the line is horizontal, as $\gamma=1$
and $1-\gamma=0$. As the correlation function reaches the regime
where only noise is left, $C(\tau)$ approaches a constant, $\gamma=0$
and $1-\gamma=1$ giving a line with slope $1$. As shown in Fig.
\ref{ttologC}, 
at short times we get a horizontal line for all temperatures, corresponding 
to the exponential decay that we have seen above. At low temperatures this 
crosses over to a straight line with some slope. 
We get straight lines for at least two orders of
magnitude for the temperatures $T={0.03-0.1}$ before the correlation function 
drops below the noise level in our data and the graphs end in a noise dominated
line of slope 1. At $T={1,10}$ we find 
straight horizontal lines that cross directly into the noise 
within our precision, without any intermediate region of stretched exponential 
behavior. 
From the lines in Fig. \ref{ttologC} we can extract three values:
the slope gives us $1-\gamma$, the offset at large $\tau$ allows us to
determine $\tau_0$, and the initial level at which the behavior is
exponential again gives us $c_T$. The fact that these initial levels
are consistently slightly below our estimates for $c_T$ reflect the
fact that locally, the width of the energy distribution is slightly
smaller than for the full simulation time.
Plots of $\gamma$ and $\tau_0$ vs
temperature are given in Fig. \ref{gammatau}.
We see that as temperature goes down, $\tau_0(T)$ increases faster
than exponential, and $\gamma(T)$ decreases almost linearly for small
$T$, in both cases with the exception of $T=0.022,0.024$. We have also
observed that the uncertainty, determined from the scattering of the
points, increases for lower temperatures, as the time before all
correlations are lost approaches the length of our time series.  For
the lowest two data points, we assume that our data are insufficient
to give correct estimates for the standard deviation, $\sigma_T$, and
mean, $\bar{E}$. Especially, if we look at too short time, we expect
to measure a too small $\sigma_T$, so the correlation functions will
therefore systematically underestimate the correlation. This will
again give too high values for $\gamma(T)$ and too low for
$\tau_0(T)$.

\begin{figure}
\includegraphics[width=8cm]{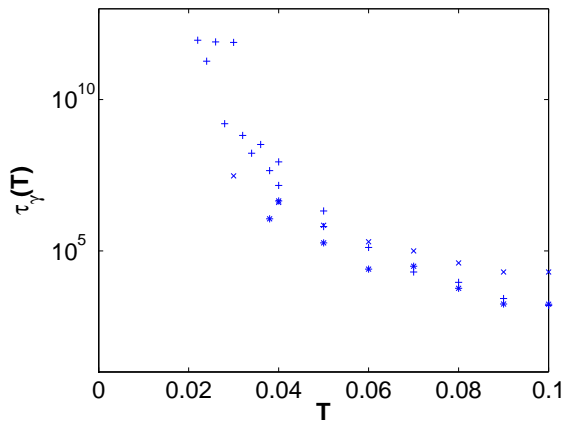}
\caption{\label{taugamma} $\tau_\gamma(T)$ from numerical integration
(*) $\tau_\gamma(T)$ from equation (\ref{taufromtau0}) (+) and the
visual estimate of the relaxation time (x).}
\end{figure}

Again following Ogielski\cite{Ogielski} we can also define a weighted time 
\begin{equation}
\tau_\gamma = \frac{\int \tau C(\tau) d\tau }{\int C(\tau) d\tau}
\end{equation}
which can be seen as a weighted average time scale for processes in the
equilibrium. Plotting $\tau_\gamma(T)$, obtained by numerical integration, 
 we see
that it increases dramatically as $T$ decreases (Fig. \ref{taugamma}). 
If we assume  $C(\tau)=e^{-(\tau/\tau_0)^\gamma}$ we can find a relation 
between $\tau_\gamma$, $\tau_0$, and $\gamma$:
\begin{equation}
\tau_\gamma = \frac{\int \tau C(\tau)d\tau}{\int C(\tau)d\tau}
 =\tau_0 \frac{\Gamma(2/\gamma)}{\Gamma(1/\gamma)} 
\label{taufromtau0}
\end{equation}
where $\Gamma$ denotes the $\Gamma$-function.  This value is also
plotted in Fig. \ref{taugamma}. The errors in these plots are huge,
especially at low temperatures, due to the following two effects.
First, due to the extreme sensitivity of the analytical estimate
Eq.(\ref{taufromtau0}) to the value of $\gamma$, any uncertainty in
$\gamma$ translates into a much larger uncertainty in
$\tau_\gamma$. Second, the long and noisy tail of $C(\tau)$ makes
numerical integration difficult. In cases where $C(\tau)$ does not
reach zero within our simulation time, numerical integration becomes
impossible. The correlation $C(t_1,t_1+\tau)$, as defined in
Eq. (\ref{corr2t}), can not be considered independent of
$C(t_2,t_2+\tau)$ unless $t_2-t_1>\tau$.  When using
Eq. (\ref{corrtau}), the number of {\it independent} time intervals we
average over can be estimated as the total simulation time divided by
$\tau$. Therefore, the noise increases with $\tau$. To improve this,
we have averaged over from 3 to 10 independent time series at the
lowest temperatures.  Still, both estimates for $\tau_\gamma$ show an
increase that is at least exponential as $T$ decreases. This again
justifies disregarding $T=\{0.024,0.026\}$ as our simulation time does
not come near these time scales.

If we assume $\gamma(T)\propto T$, which does not seem impossible for
small $T$, the gamma functions in Eq.\ref{taufromtau0} can be
estimated using Stirling's formula to give an estimate for
$\tau_\gamma$ as
\begin{equation}
  \tau_\gamma \propto \tau_0(T) 
  \frac{1}{T}^{\frac{1}{T}}e^{-\frac{1}{T}}2^{\frac{1}{T}}
\end{equation}
There is no way we can differentiate between this kind of behavior
and a divergence at a specific temperature within our data. Any divergence 
in $\tau_{\gamma}$ would have to come from either $\gamma$ going to zero, 
which appears to happen at $T=0$ if we extrapolate linearly, or from 
a divergence in $\tau_0$, which we have not been able to identify.

\subsection{Global density of states}\label{dos}

Originally, the dynamics of the Coulomb Glass were understood mainly
from the single electron jump picture, giving rise to the theory of
the Coulomb Gap,\cite{EfrosShklovskii} which is a gap in the single
electron density of states. But in order to say something about
behavior at higher temperatures, we should also know something about
the global density of states far away from the ground state,
especially as our results show that all our visited states are likely
to be far from the ground state.

The probability $P(E)$ of the system being at an energy $E$ is assumed
to be simply the product of the probability of a given state being
occupied times the density of states at this energy, $g(E)$, giving
\begin{equation}
P(E) = \frac{1}{Z} g(E) f(E,T), \qquad Z = \sum_{E'} f(E',T) g(E')
\label{defPE}
\end{equation}
where $f(E,T)=e^{-\beta E}$, $\beta=1/T$, as we have chosen to measure
temperature in the same units as energy. If we assume that the simulation
time suffices for the distribution of energies we
observe to be representative for the behavior at infinite times, we can
use the observed $P(E)$ to estimate $g(E)$. If we furthermore
assume the system to be ergodic, $g(E)$ will be the actual
density of states of the full system. If not, $g(E)$ is just the
density of states accessible from the set of states we have visited,
while the global density of states is a sum of different $g(E)$-s. We
have tried relaxing the system from different electron configurations,
and seen no trace of variation in the equilibrium distribution. While
this does not constitute any proof, we can at least conclude that the
system is either ergodic, or that $g(E)$ is the same for multiple
separate sets of configurations. 

\begin{figure}
\begin{tabular}{ccc}
\includegraphics[width=5cm]{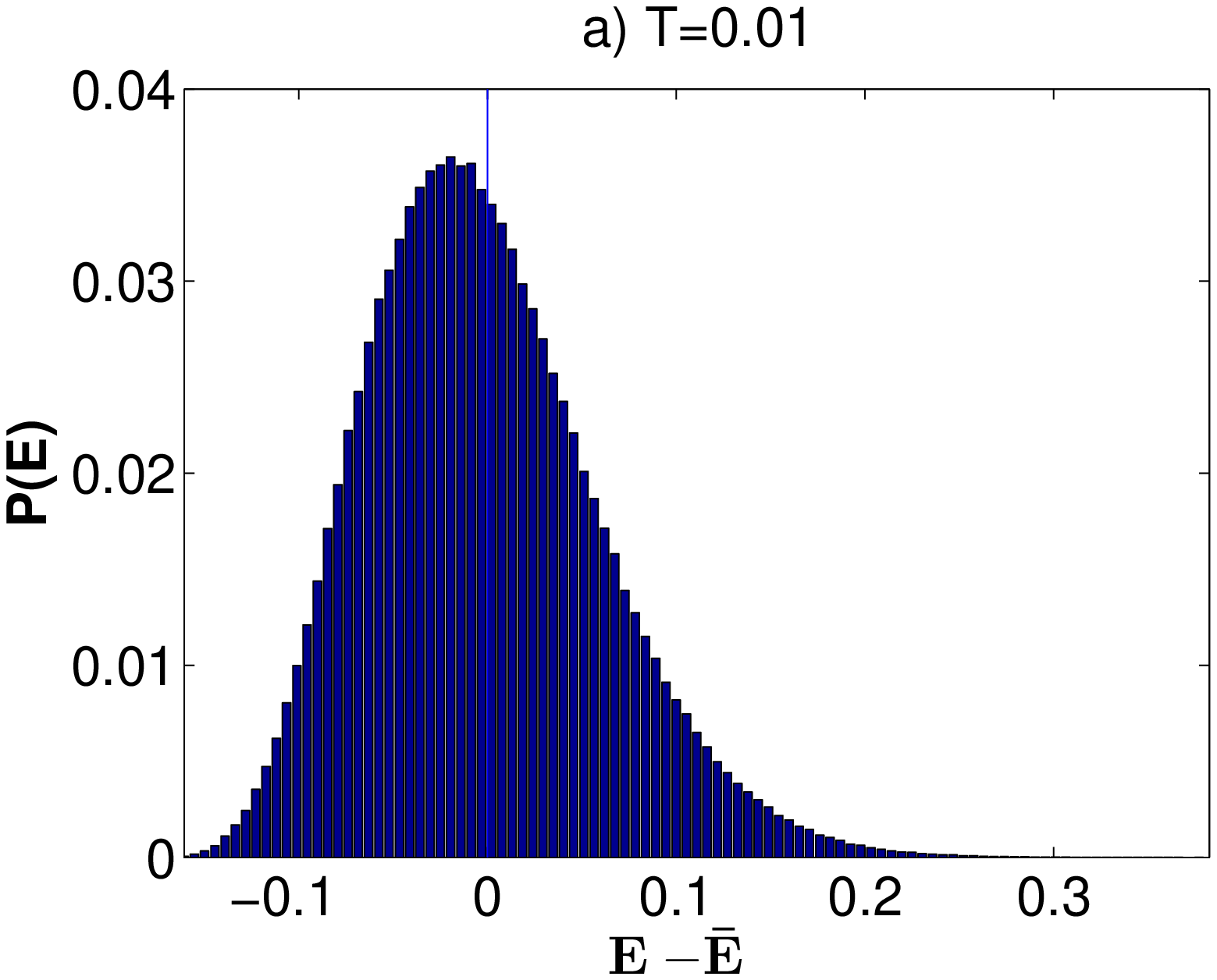}&
\includegraphics[width=5cm]{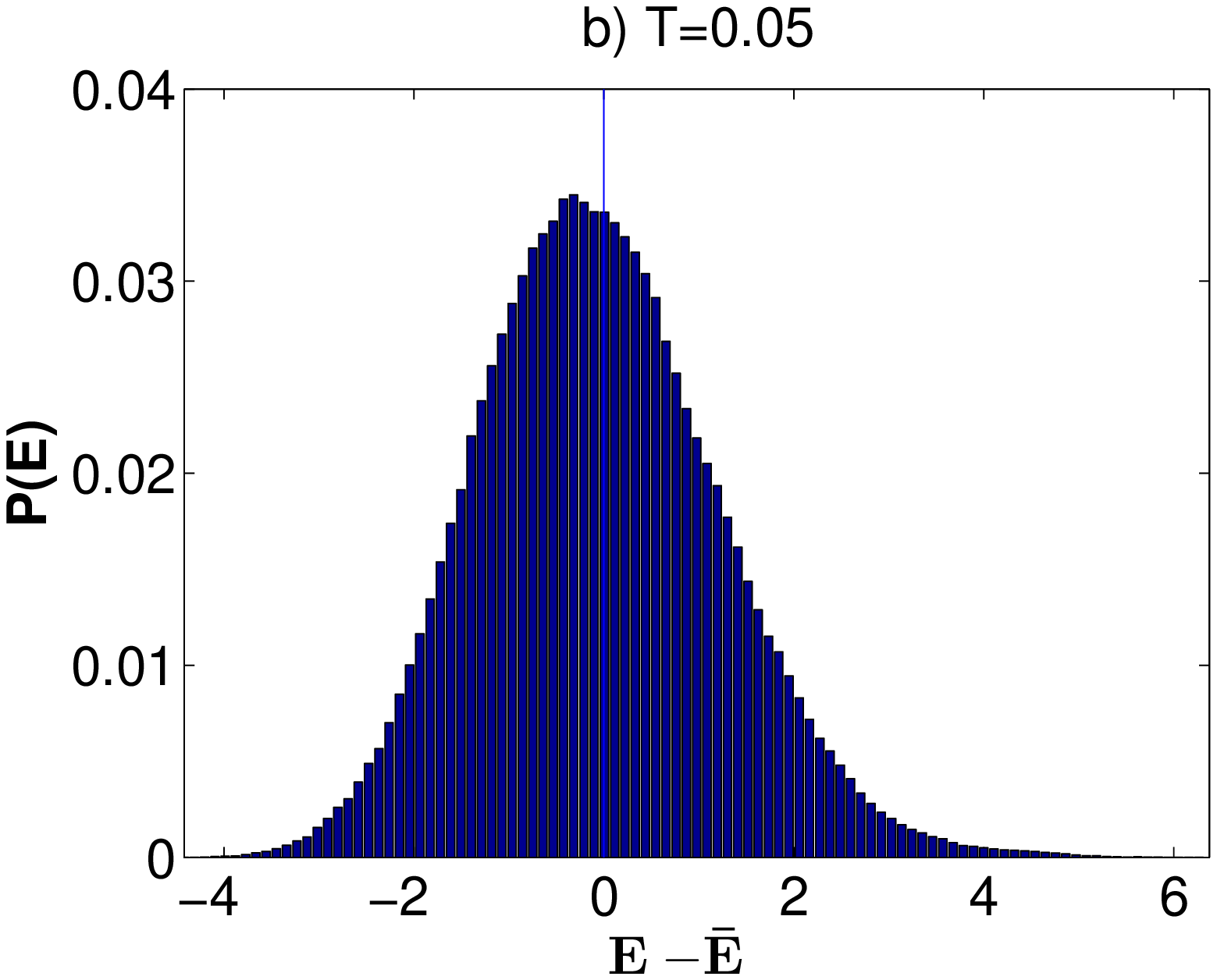}&
\includegraphics[width=5cm]{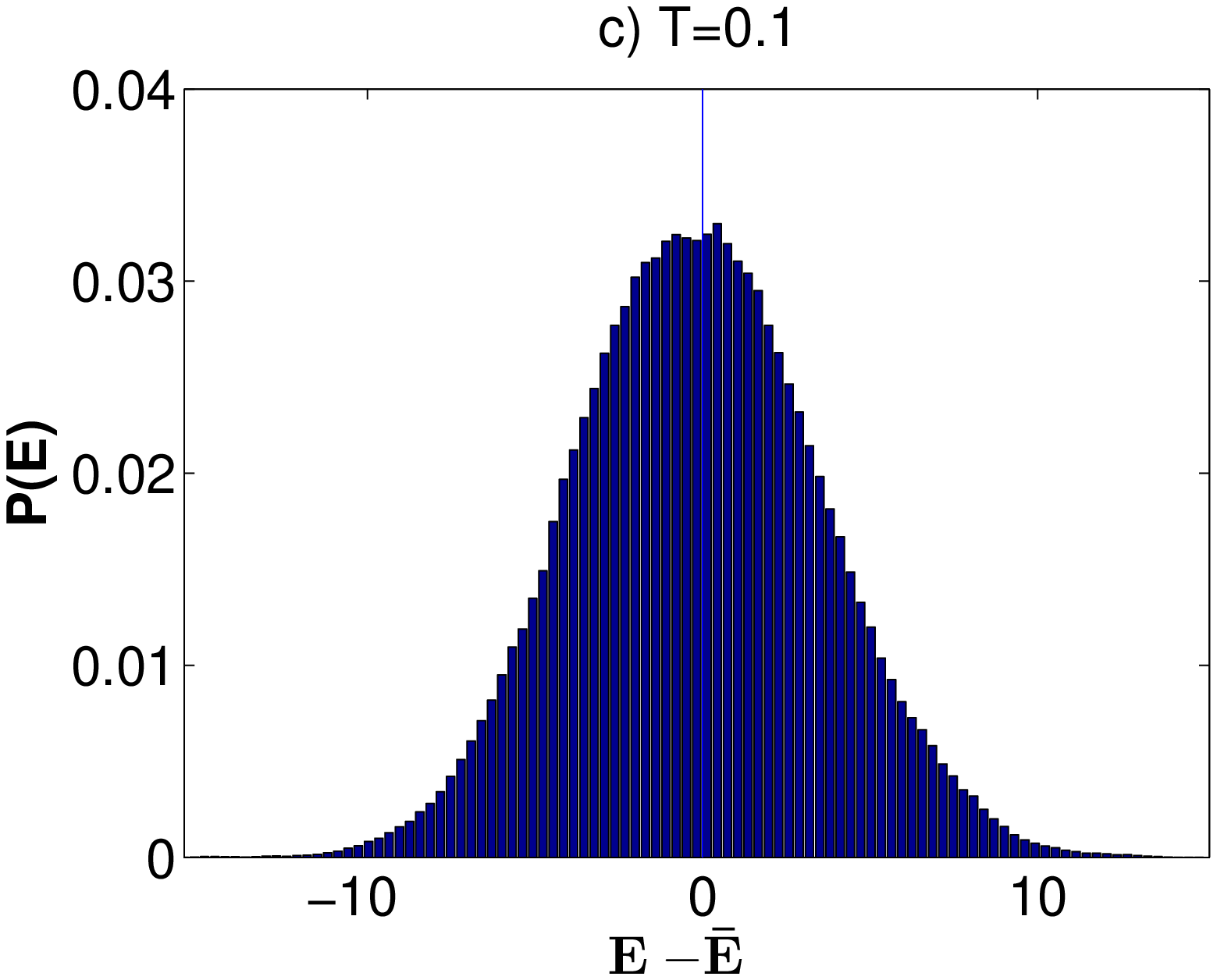}
\end{tabular}
\caption{\label{PE} Histograms of $P(E)$ for T=\{0.01,0.05,0.1\}. The
  histogram is for the last $10^6$ jumps of a run of $10^7$ jumps,
  except for $T=0.01$, where only $8\cdot10^5$ where included to stay
  within one step. It is seen that at high temperature the
  distribution is symmetric, whereas the lack of symmetry for $T=0.01$
  is clear. The vertical lines indicate $\bar{E}$, the mean of the
  equilibrium distribution.}
\end{figure}

\begin{figure}
\begin{tabular}{ccc}
\includegraphics[width=5cm]{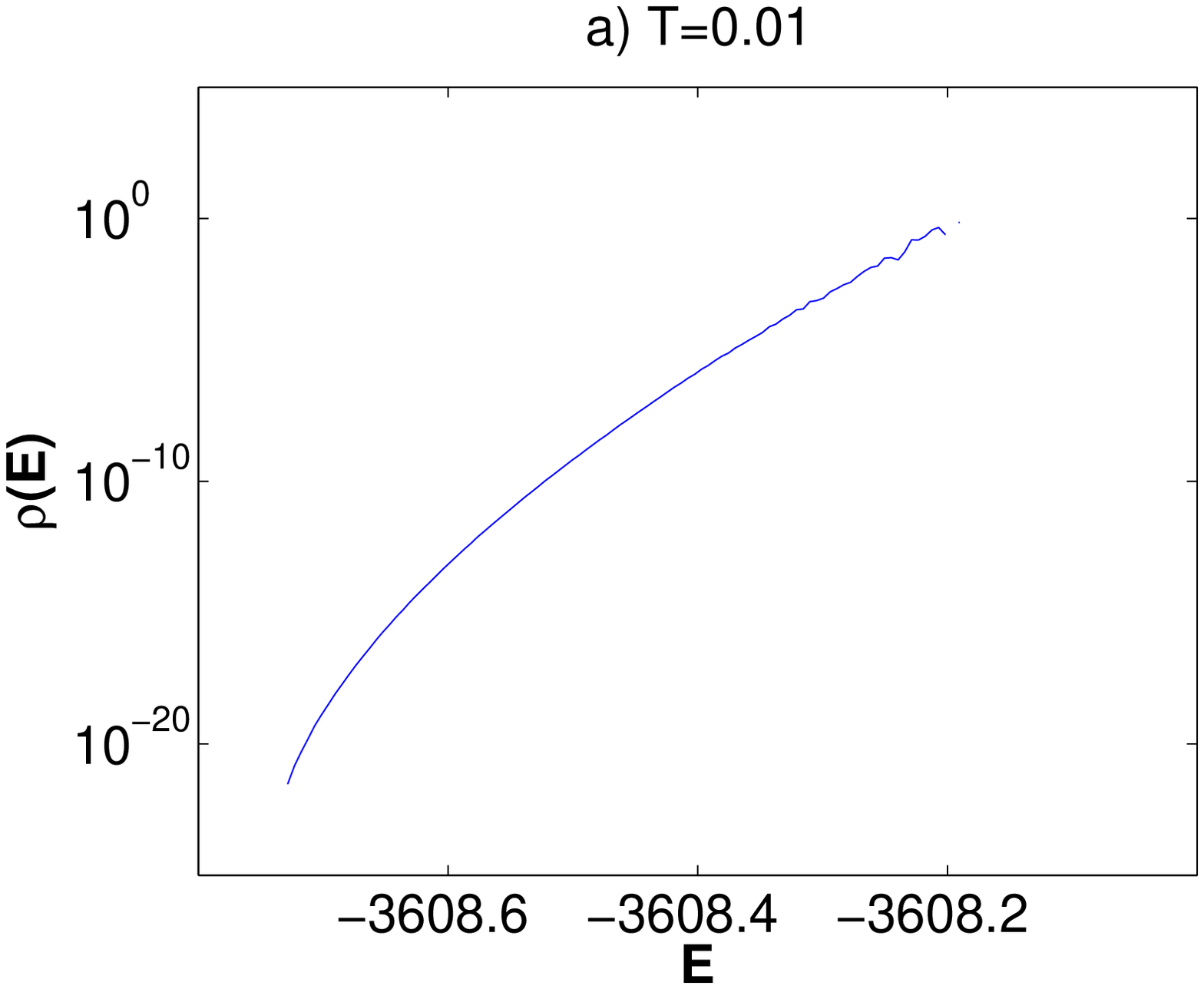}&
\includegraphics[width=5cm]{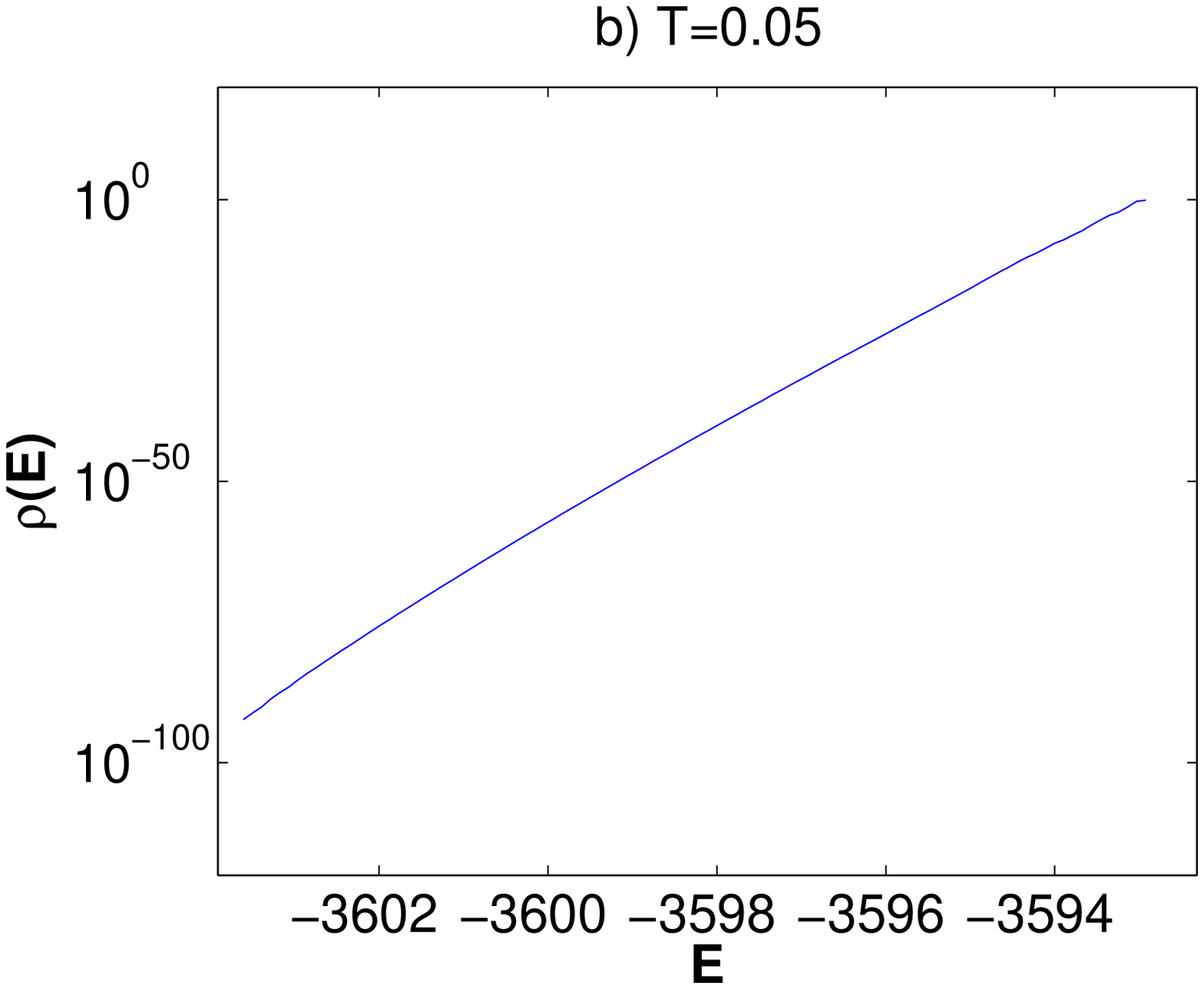}&
\includegraphics[width=5cm]{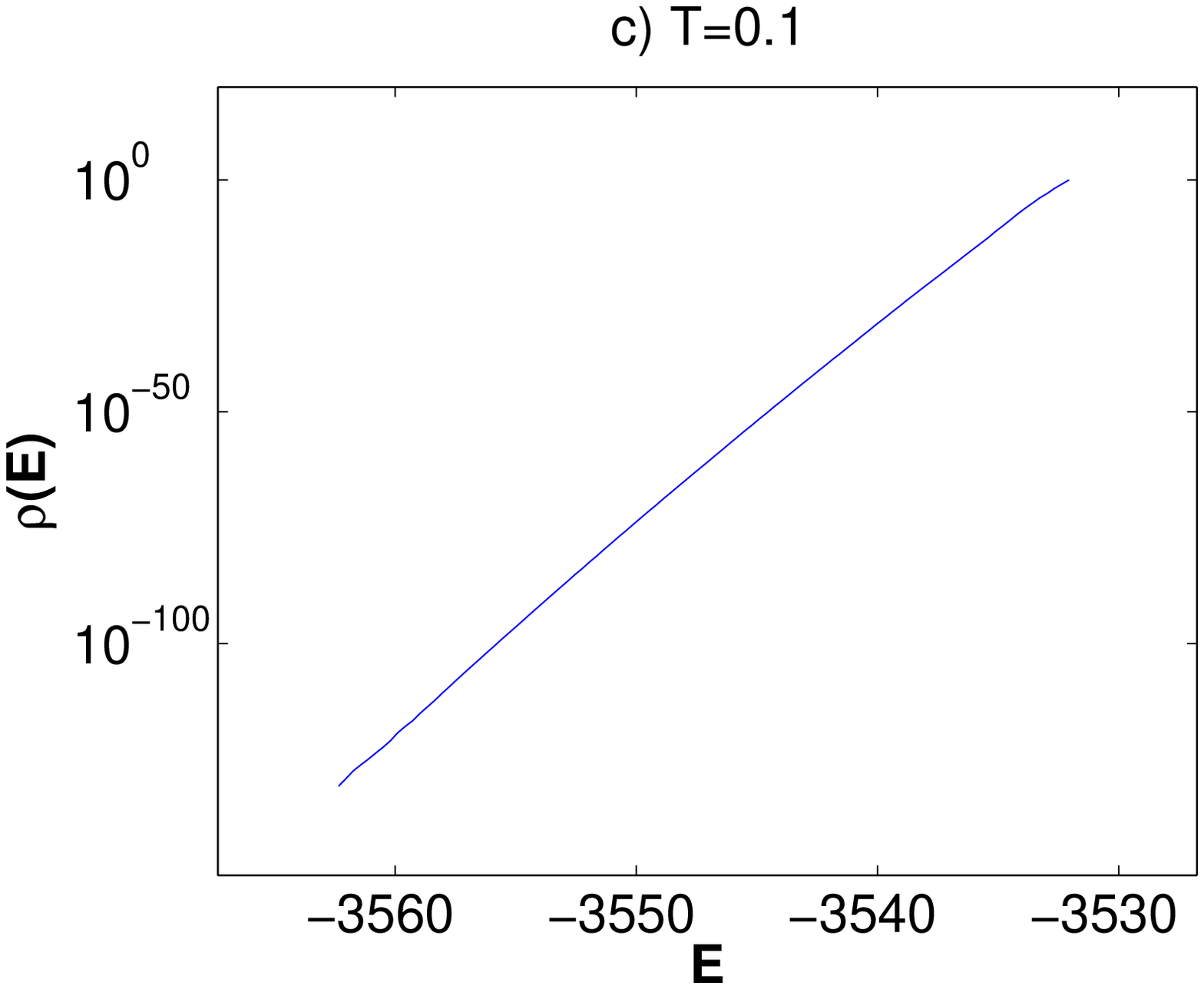}
\end{tabular}
\caption{\label{gE}Estimated $g(E)/g(E_T)$ from the distribution at
  $T=\{0.01,0.05,0.1\}$. For $T=0.01$ we clearly see that
$\rho=\ln(g(E))$ has a strong curvature, for $T=\{0.05,0.1\}$ the
curvature is less pronounced, but still present.}
\end{figure}

At low temperatures, where we are not able to reach equilibrium we can
still consider one step (as defined in Fig. \ref{zoomrel}b) on a descending graph. Then we get the
density of states for the set of states accessible by likely
transitions, a set sometimes referred to as a valley in configuration
space. This is the case in the plots for $T=0.01$ in Figs. \ref{PE}
and \ref{gE}.

Fig. \ref{PE} shows  $P(E)$ 
calculated from the number of times a state with energy $E$ is
visited, weighted by the time spent in that configuration.

Solving Eq. (\ref{defPE}) for $g(E)$ we write
 \begin{equation}
g(E) = P(E) f(E,T)^{-1} \sum_{E'} f(E',T) g(E') 
\end{equation}
Note that $g(E)$ is written as a vector of values for discrete
energies $E$. This is due to our discretization of $P(E)$ in the form of
histogram boxes. Defining the matrix $M_{EE'}=P(E) f(E,T)
\sum_{E'} f(E',T)$ we can write this as
\begin{equation}
g(E) =\sum_{E'} M_{EE'} g(E')
\end{equation}
Thus $g(E)$ must be an eigenvector of the matrix $M_{EE'}$, with
eigenvalue 1. In this way we can find $g(E)$ except for a constant
prefactor $g_0$.

The calculated $g(E)/g_0$ for our equilibrium distributions
are shown in Fig. \ref{gE}.
Note that the plot shows $\ln (g(E)/g(E_T))$, where $E_T$ is the
highest energy obtained at equilibrium for this temperature. Within
the energy region covered by each run, the density of states
approaches a straight line, for more than hundred orders of magnitude
in $g$ in the case of $T=0.1$.

Based on the obtained results, we choose to write $g(E)=g_0 e^{\rho(E)}$,
where $\rho(E)$ is well defined for energies well above the ground state,
$E_0$.
\begin{equation}
P(E)=\frac{e^{\rho(E)-\beta E}}{Z}
\end{equation}
has a maximum when
\begin{equation}
\frac{d P(E)}{dE}=\frac{1}{Z}\left(\frac{d\rho(E)}{dE}-\beta\right)
e^{\rho(E)-\beta E}
\end{equation}
equals zero.
We can thus expect the maximum of the distribution, $E_m$, to be in an area
where the rise in DOS is comparable to the inverse 
temperature, $\rho'(E_m)=\beta$.
Lower temperature requires a steep rise in DOS to reach equilibrium, 
whereas at higher
temperature equilibration will occur with slower increases in DOS. 

To find the width of the distribution we write $E=E_m+\Delta E$, where
$E_m$ is the previously obtained energy corresponding to the maximum
of the distribution, for which we know that $\rho'(E_m)=\beta$. An
expansion of $P(E)$ to the second order in $\Delta E$ then gives:
\begin{equation}
P(E)=\frac{1}{Z} e^{\rho(E)-\beta E}
 =\frac{1}{Z}e^{\rho(E_m)+\rho'(E_m) \Delta E 
 + \rho''(E_m)\Delta E^2/2-\beta (E_m+\Delta E)}.
\end{equation} 
Separating the $E_m$ dependence from the $\Delta E$ dependence this gives
\begin{equation}
P(E) =\frac{e^{\rho(E_m)-\beta E_m}}{Z}e^{\rho''(E_m)\Delta E^2/2},
\end{equation}
which shows that the standard deviation of the distribution is $\sigma
= 1/\sqrt{|\rho''(E_m)|}$. In this way we can readily find both
$\rho'(E)$ and $\rho''(E)$ at as many energies as the number of
temperatures at which we run relaxations.

The most immediate use of this information is probably for algorithms
requiring the mapping of all states accessible for the
system.\cite{Pollak} Estimating $P(E)$ gives an upper temperature
limit for which one can hope to map all states, for a given system
size.

\section{Discussion}

\subsection{Validity of the random walk model}
We see that for high temperatures, $T=\{1,10\}$ the model of the
random walker seems to be a good description of the observed dynamics.
Regarding both the density of states and the short time correlations,
the same is true for the lowest temperatures still reaching an
equilibrium. But for longer times, the correlations persist much
longer than our estimates for $c_T$ should indicate.

Some general understanding that has been suggested\cite{Campbell} is
that at high temperatures the network of thermally allowed transitions
is sufficiently dense in the space of configurations that the number
of possible jumps is a function of the density of states (DOS)
only. The system performs a random walk on this network, which when
projected on the energy gives the random walk in energy that we have
considered above, and exponential decay of the correlation
function. Below some temperature $T_c$ the network of allowed
transitions becomes diluted and has some fractal structure. A random
walk on this fractal leads to anomalous diffusion
$\left<r^2\right>\propto t^\gamma$, which is believed to correspond to
a stretched exponential $e^{-(\tau/\tau_0)^\gamma}$ for the
correlation function of any confined quantity like our energy, both
with the same exponent $\gamma$ (As far as we are aware, this has only
been numerically confirmed for hypercubes\cite{Campbell} and
hyperspheres\cite{Jund} and not rigorously established). Thus, we
expect that as temperature increases, $\gamma$ should also increase,
and approach 1 at $T_c$. We see from Fig.  \ref{gammatau} that this
seems to be the case, but our data are not sufficient to determine
$T_c$. In this temperature regime a standard Metropolis algorithm
would be more suitable than our low temperature algorithm.  It has
also been suggested\cite{Campbell} that because the configuration
space has a large dimension, $\gamma$ should approach the mean field
value of percolation, $\frac{1}{3}$, at the glass transition. This was
indeed the case in the work of Ogielski,\cite{Ogielski} but is clearly
not the case in our simulations. $\gamma$ reaches a value of
approximately 0.15 at $T=0.03$, but can possibly be estimated to even
lower values with better data sets.

\subsection{Possible model system}

To give a picture of what kind of model could adequately describe this
picture, we start from a landscape of local minima, defined by the
fact that there exist no single electron transitions taking the system
down in energy. We have previously shown that there exist a huge
number of such minima.\cite{Glatz}

\begin{figure} 
\includegraphics[width=10cm]{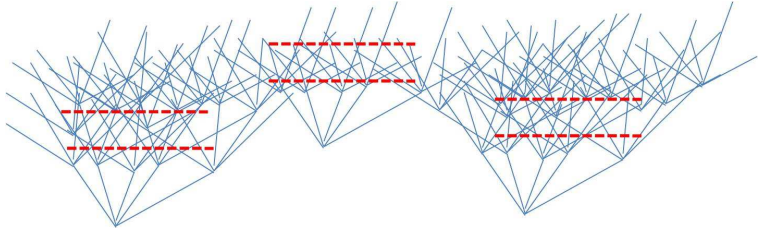}
\caption{\label{CGstates} Simplified picture of the configuration space. The dotted lines indicate regions with similar increase in density of states, but at different energies, due to different starting points.}
\end{figure}

From each local minimum, we can find all states that can be reached by
one single electron jump, by definition taking the system up in
energy.  From each of these states, we can again add states accessible
by another jump, and so on, but keeping only those that increase the
energy.  In this way we create a tree growing up from each local
minimum.  It is easy to imagine these single trees having an initial
exponential growth in the density of states. As the states of one tree
become identified with states of other trees, the exponential growth
slows because the number of identified states increases
drastically. We have attempted to illustrate this in
Fig. \ref{CGstates}
The branch thickness is described by the wave function overlap of the
two states involved in that specific transition, and is temperature
independent. In addition, at each temperature, there is a likelihood
of passing up the branch different from that of passing down the
branch, given by energy difference, temperature, and branch thickness.
If we define a minimum jump probability given by the time scale of our
experiment, we can cut away all branches with lower probabilities than
the limiting value, like in a random resistor network. As temperature
goes down, more and more branches will be cut or made one-way streets
for the purpose of system development. The resulting network will
resemble a fractal network with a temperature dependent dimension. 

A random walker on such a network will show dynamics on many competing
time scales. First, within the branches of one 'tree', there is the
time scale to make single jumps, corresponding to our $\tau_T$. Then
comes the time scale for getting from the bottom of a 'tree' to the
top, or opposite, which is approximately $1/c_T$. Then there is the
probability of jumping from one 'tree' to another, from one cluster of
'trees' to another, between clusters of clusters and so on. The
average energy of the clusters can vary, giving correlations over long
time scales as we slowly climb from one cluster to the next. This
process is described by the slow decay of the correlation function
$C(\tau)$.  When temperature is lowered, the probability of reaching
the top branches will be reduced, and there will be fewer or less
likely connections between the different 'trees' and clusters. The
probability of jumping to lower clusters will decrease, but that of
jumping to higher energy clusters will decrease even more. Such a
random walk should behave much like our simulation.

While we believe that such a model will lead to the observed behavior,
we have no independent verification that this model actually corresponds
to our system.

\subsection{Temperature and time scales}

As mentioned in the introduction, one of our objectives in 
undertaking this project was to identify a glass transition 
temperature and a dynamic phase diagram for the lattice model. 
While we have not been able to identify a transition temperature, 
or determine whether such a finite temperature exists, we have 
identified timescales where the system will equilibrate as 
function of temperature. If we wish to study the time evolution of 
some quantity, we have to average over time intervals shorter than 
this equilibration time. 

Tsigankov et al.\cite{Tsigankov} studied the lattice model that we
have used and compared it with the random site model. They were interested
in the long time relaxation of the conductance observed in
experiments,\cite{Ovadyahu} but since the direct simulation of this
process would take too long time, they calculated the conductance and
the shape of the Coulomb gap for various initial states, arguing that
if one is to observe slow relaxation one would need to find metastable
states with sufficient spread in conductance. Their simulations were
performed at $T=0.04$ as this was the lowest temperature they were
able to use for their algorithm to be efficient. They observed that
the variation in the conductance for the lattice model was too small
to explain the observed change in the conductance during slow
relaxation and concluded that the lattice model can not be used to
explain the experiments. With our optimized algorithm we are able to
go below this and we have shown that we are able to reach equilibrium
for temperatures at least down to $T=0.03$, possibly as low as
$T=0.02$. Therefore we believe that the results of Tsigankov et
al.\cite{Tsigankov} are probably a result of using a temperature where
the system equilibrates, and that their conclusion could be different
if they could repeat their simulation at lower temperature, below some
glass transition temperature $T_g$, or at least in a region where
$\tau_\gamma$ is greater than the simulation time span. It should be
noted that Tsigankov et. al. use localization length $a_l=1$ rather
than our $a_l=2/3$. We believe that increasing $a_l$ allows more
connections between the different regions of the configuration space,
in the same way as an increase of temperature. Thus, Tsigankov's
temperature $T=0.04$ would be equivalent to a higher temperature in
our simulations. This argument agrees well with
the relation between temperature and localization length in the
expression for conductance, which is a function of the product
$a_lT$.\cite{EfrosShklovskii,Tsigankov2}

\section{Conclusions}

Based on the above analysis, we have reached the following
conclusions:
\begin{itemize}
\item{For temperatures down to a limiting $T_{min} \approx 0.02$ we
have demonstrated that the system equilibrates within our simulation
time. This equilibration occurs at a total system energy for which the
total number of accessible states is so high that a full mapping of
the states is only possible for very small systems.}
\item{At temperatures close to, but above $T_{min}$ we observe energy
correlations following a stretched exponential law. The exponent of
this stretched exponential, $\gamma$, decreases with temperature,
seemingly with a linear dependence for low temperatures. At
temperatures $T\gtrsim 1$, $\gamma$ approaches unity, but determining
the exact behavior requires further study.}
\item{$\tau_\gamma$, the average relaxation time weighted by the
correlation function, increases rapidly with decreasing
temperature. It seems to roughly follow the time needed to establish
equilibrium.}
\item{The rapid increase of $\tau_{\gamma}$ with decreasing temperature,
means that the low temperature limit for when equilibrium can be
established is only weakly dependent on the total simulation
time. From our data we cannot conclude whether $\tau_{\gamma}$ actually
diverges at any finite temperature, or whether it can be used to
define a glass transition.}
\item{The observed behavior is compatible with a model of the system
as a random walk on a fractal configuration space.}
\item{There exists a temperature range for which the lattice model
with single electron hops only, can probably be used to study slow
dynamics. }
\end{itemize}

It is important to stress that we have shown the time scales of the
single jump dynamics only. As the correlation times become longer,
processes that are unlikely when looking at single changes of state
may still be important for the long term dynamics of the system. We
have only shown that single jumps can give long time scales, not that
these time scales will actually be present in the model if other
dynamics are also allowed. Specifically, multi-electron jumps are
likely to hasten the transitions between state clusters at low
temperatures, giving shorter correlation times and larger values for
$\gamma(T)$.

\acknowledgments
The code used in this program is based on code generously made
available to us by Andreas Glatz at Argonne National Laboratories.
The project has been financed by the STORFORSK program of the
Norwegian Research Council. We also thank Yu. M. Galperin and A. Voje for
critical reading of the manuscript.

\appendix
\section{Details of relaxation algorithm and parameter choices}
\label{algorithm}

We use Monte Carlo simulation, first calculating the probabilities of
all those jumps that are likely to occur from a given configuration,
and choosing one of them to actually happen. The physical time
spent on one such step is calculated as the inverse of the sum of all
the rates for individual jumps. The limitation of the basic model is
therefore that it allows single particle jumps only.

The rate for a single jump is given in Eq. (\ref{tunnelingrate}). This
expression is only strictly valid when there is a constant barrier
height. In our system this is not the case, as the charge distribution
gives valleys and peaks in the barrier over the volume the electron
wave function covers. A proper treatment would have to include an
integral over all possible paths. We still follow the tradition in the
field and use the expression above, as the exact alternative would be
impossible to implement.

Calculating all possible jumps is still a very time-consuming process.
Following Ortu\~no et al. \cite{Ortuno} we therefore limit ourselves
to calculating those jumps that are probable to occur. We see that eq
(\ref{tunnelingrate}) decays exponentially both with distance and
energy difference. Therefore, a very few rates will be big, while the
majority of rates will be neglectably small. It is therefore possible
to write
\begin{equation}
\Gamma_{tot}=\sum_{i,j}\Gamma_{i \rightarrow j} =
\sum_{prob}\Gamma_{i \rightarrow j}+ \sum_{improb}\Gamma_{i\rightarrow j}
 \approx \sum_{prob}\Gamma_{i \rightarrow j}
\end{equation}
assuming the latter sum, over improbable jumps, to be negligibly
small. The first sum, the probable jumps, can be shown to include only
a relatively small number of jumps, depending on the configuration. 
A similar thought is used in the algorithm presented
by Matulewski et. al.\cite{Matulewski2}, even though their
implementation is more static than ours.

\begin{figure}
\includegraphics[width=8cm]{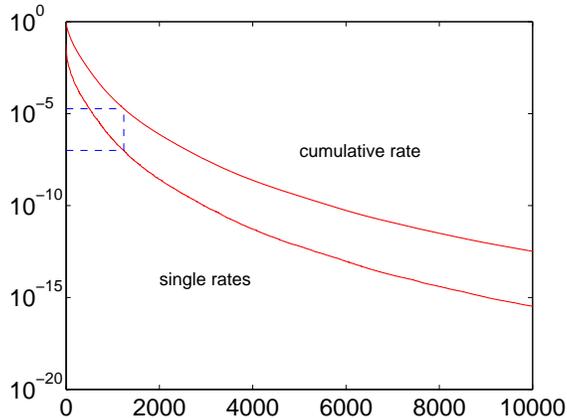}
\caption{\label{test208} $\Gamma_n/\Gamma_{tot}$ and
  $(\sum_{n'<n}\Gamma_{n'})/\Gamma_{tot}$ vs $n$. Dotted lines
  indicate the cut-off limit and corresponding error.}
\end{figure}

We define a
limiting rate $\Gamma_{min}$ so that only jumps where $\Gamma_{i
  \rightarrow j} > \Gamma_{min}$ are included in the first sum.
\begin{equation}
\Gamma_{min}= \Gamma_0 e^{-M}
\end{equation}
where $M$ is the maximal allowable exponent to give rates higher than
$\Gamma_{min}$. For the simulations presented we have chosen to set
$\Gamma_{min}$ to $10^{-7} \Gamma_{tot}$ for the higher temperatures,
where our algorithm is slow, and $10^{-10} \Gamma_{tot}$ for low
temperatures. Since $\Gamma_{tot}$ is only known after all the rates
are calculated, we use the $\Gamma_{tot}$ from the previous step in
calculating $\Gamma_{min}$.  The validity of our approach can be
tested by varying the ratio $\Gamma_{min}/\Gamma_{tot}$, and see
whether it influences the dynamics, as shown in Fig. \ref{ttologC}. A
simple estimate on the error made can be obtained as illustrated in
Fig.  \ref{test208}. Here we have calculated the rates of all jumps,
and sorted them by magnitude. We define $\Gamma_n$ as the $n$'th
largest rate. Plotting the individual and cumulative probabilities
together, we can immediately read off the error we make if we cut off
all rates with a value less than a certain fraction of the total
rate. Our cut-off at $10^{-7}\Gamma_{tot}$ means that in this randomly
chosen instance at $T=0.02$, we would need to include 1200 jumps, and
make an error in the total rate of less than $2\cdot
10^{-5}\Gamma_{tot}$.  This means that approximately 10 jumps are
erroneously cut off in $10^6$ steps.
The selection of which rates to calculate is optimized using the $M$
defined above. We wish to cut off improbable rates, that is rates
where the increase in energy is much larger than temperature, or where
distances are very long. From the expression given for the tunneling
rate in Eq. \ref{tunnelingrate}, using the expression for phonon
absorption, we can take the $1$ in the denominator to be small, which
simplifies our requirement for the limiting tunneling rate to the
expression
\begin{equation} 
|\Delta E_{i \rightarrow j}|/T + 2 r_{i,j}/a_l < M,
\end{equation}
disregarding the preexponential factor which further reduces the rate
for low temperatures. We also know that the energy
difference is given as
\begin{equation}
\Delta E_{i \rightarrow j}= \epsilon_j-\epsilon_i-1/r_{i,j}
\end{equation}
and the minimum value for $\epsilon_j$, $\epsilon_{j,min}$ can easily
be found from the list of single particle energies. So
\begin{equation}
\epsilon_{j,min}-\epsilon_i-\frac{1}{r_{i,j}} \le 
\epsilon_{j}-\epsilon_i-\frac{1}{r_{i,j}} =
\Delta E_{i\rightarrow j}\le  T \left(M - \frac{2 r_{i,j}}{a_l}\right)
\end{equation}
Thus for each $\epsilon_i$ there is a given maximal radius that can
potentially give probable jumps. Conversely, for each radius, there is
a lowest allowable $\epsilon_i(r)$. If $\epsilon_i(r)$ is below this
limit, there is no point in looking for empty sites further away than
the given radius. 

In the initial stages, many jumps are very likely, so $\Gamma_{tot}$
and therefore also $\Gamma_{min}$ is very large. This gives a small
maximal radius for all sites, regardless of $\epsilon_i$. While long
jumps could possibly happen, it is much more likely that a short jump
happens first.

As soon as the Coulomb Gap has formed, there will be very few sites
where the $\epsilon_i$ is sufficiently high to allow long jumps to be
probable. Only for these occupied sites do we have to check many
possible destinations, and many sites will have no likely jumps at
all. Thus for both situations, with and without gap, the number of
calculated rates scale roughly as $N$ rather than $N^2$.

As observed by Ortu\~no et al.,\cite{Ortuno} there is often a limited
number of jumps that are repeated a very large number of times.
Ortu\~no et al.\cite{Ortuno} solve this by calculating the rate to
escape from an ensemble of configurations. We have used a faster and
easier approach, where we simply save the state and jump rates for
each configuration, until a jump occurs that is unlikely to be
reversed. Every time a state is revisited, we only have to switch some
pointers. The disadvantage of this approach is that we record a huge
number of jumps that are simply repetitions, but it does not cost us
much computation time. The advantage is that we record the time spent
at all configurations explicitly, making it possible to easily extract
information on thermal properties, noise etc. Also, our algorithm is
faster in those cases where a configuration is not revisited. For very
low temperatures, $T\lesssim 0.001$, the number of repeated jumps
becomes impracticable large. In this case Ortu\~no's algorithm will
probably work better than ours.

The algorithm used is optimized to work best for temperatures where
the Coulomb Gap has formed properly, so both the energy and the
distance terms give fast convergence for the sums. Thus the initial
phase of relaxation will be computationally slower per step. Also,
higher temperatures give a slower convergence in the energy term, and
thus longer processing times. At some temperature, the optimized 
Metropolis algorithm of Ref. \onlinecite{Tsigankov} becomes significantly more 
efficient. Preliminary tests indicate that this happens around
$T=0.05$.

In order to have a constant ground state and density of states, all
simulations presented in this article are based on the same
realization of disorder of the lattice. Some preliminary results for
conductance simulations indicate that a system of $100^2$ sites may
still be too small to avoid size effects for diffusive processes. On
the other hand, we expect a requirement for longer time series for
bigger systems, and the combined increase in computational effort of
the bigger system and longer time makes it impractical to simulate.
For most of the simulations we have been limited to using desktop
computers, reducing our capacity for obtaining optimal statistics, but
we still consider the results to be sufficient to give trustworthy
conclusions.

The localization length is somewhat arbitrarily set to $2/3 a$, where
$a$ is the lattice constant. If we use localization lengths larger
than $a$, mixing of the states would give significant changes in the
wave functions. If we use a very short localization length, jumps
longer than to the nearest neighbor will become highly improbable at
all but the very lowest temperatures, enhancing the importance of the
quadratic lattice. $a_l=2/3 a$ gives a significant number of jumps at
distances up to three sites away, but rarely longer, for the most
relevant temperature range. 

\section{Theory of a Random Walker in a potential}
\label{randomwalker}

Let $u(x,t)$ be the probability of finding the random walker with
position $x$ at time $t$. Let it take steps of length $\delta$ either
increasing or decreasing its energy and let $\tau$ be the time of each step. 
Then the master equation for $u(x,t)$ is 

\[
 u(x,t+\tau) = p(x-\delta) u(x-\delta,t)+ q(x+\delta) u(x+\delta,t)
\]
where $p(x)$ is the probability of making a step in the direction of 
increasing $x$ when the walker is at energy $x$ and $q(x)$ the 
probability of making a decreasing step. Expanding to first order in 
$\tau$ and second order in $\delta$ we get 

\begin{equation}\label{master}
 \frac{\partial u}{\partial t} = 
  (q-p)\frac{\delta}{\tau}\frac{\partial u}{\partial x}
 + \frac{\partial }{\partial x}(q-p)\frac{\delta}{\tau}u
 + \frac{\delta^2}{2\tau}\frac{\partial^2 u}{\partial x^2}
\end{equation}
where we have used that 
\begin{equation}
\frac{\partial}{\partial x}(p+q)=\frac{\partial^2}{\partial x^2}(p+q)=0.
\end{equation}
We need to find expressions for $p(x)$ and $q(x)$. The requirement 
of microscopic balance (and one can check that our Monte Carlo algorithm 
satisfies this) is

\begin{equation}\label{microbalance}
 p(x) P(x) = q(x+\delta) P(x+\delta)
\end{equation}
where $P(x)$ is the probability of finding the system at position $x$. As
discussed in Sec. \ref{dos} this can be written
$P(x)=Ae^{\frac{\rho''}{2}x^2}$ close to the equilibrium position.  We
will assume that the probability $q(x)$ changes slowly on the scale of
a single step $\delta$. This allows us to replace $q(x+\delta)$ with
$q(x)$ on the right hand side of Eq. (\ref{microbalance}) and 
using $p(x)+q(x)=1$ we get 

\[
  p(x) = \frac{e^{\rho''\delta x}}{1+e^{\rho''\delta x}}, \quad 
  q(x) = \frac{1}{1+e^{\rho''\delta x}},
\]
where we have omitted the $\delta^2$-term in the exponent.
Expanding to lowest order in $\delta$ we find
\[
q(x)-p(x) = -\frac{1}{2}\rho''\delta x
\]
Inserting this in Eq. (\ref{master}) we get
\begin{equation}\label{diffusion}
\frac{\partial u(x,t)}{\partial t} = D \frac{\partial^2 u(x,t)}{\partial x^2}+c \frac{\partial}{\partial x}(x u(x,T))
\end{equation}
where $D= \frac{\delta^2}{2\tau}$ and $c=\frac{\delta^2}{2\sigma^2\tau}$. 
Eq. (\ref{diffusion}) is the diffusion equation in the presence of a 
harmonic potential. 
This is identical to the problem of momentum distribution of a
particle under Brownian motion, and an exact solution has been provided by
Chandrasekar\cite{Chandrasekar}: 
\begin{equation}
u(x,x_0,t) = \frac{1}{\sqrt{2 \pi}} \sqrt{\frac{c}{D(1-e^{-2ct})}}e^{-\frac{(x-x_0 e^{-ct})^2c}{2D(1-e^{-2ct})}}.
\end{equation}
where $x_0$ is the starting position. We see that this solution has
the properties we expect, becoming a delta function for $t \rightarrow
0$, it always has a Gaussian shape, and as $t \rightarrow \infty$ we
get standard deviation $\sigma=\sqrt{D/c}$ and mean $0$ as expected.

Looking at the expectation value of the position as function of time, 
we find
\begin{equation}
\left< x(t) \right> = x_0 e^{-ct}.
\end{equation}
Thus if we initiate the system at at position $x_0$, we expect it to
relax exponentially towards the equilibrium value as $e^{-ct}$.

We can now define the probabilities $P(E_0)$, the probability of
starting at a given energy $E_0$, and $P(E,t|E_0,0)$, the probability
of being at $E$ at time $t$ given that the system was at $E_0$ at time
$0$. From the two-time energy correlation function we can identify
$P(E_0)=u(t=\infty)$, while $P(E,t|E_0,0)=u(E,E_0,t)$, giving
\begin{equation}
C(t)=\frac{1}{\sigma^2}\int\int d E_0 dE\, E_0 E u(\infty)u(E,E_0,t)
\end{equation}
\begin{equation}
=\frac{1}{2 \pi\sigma^2} \frac{c}{D}\frac{1}{\sqrt{1-e^{-2ct}}}\int \int dE_0 dE E_0 E
e^{-\left[ \frac{cE_0^2}{2D}+\frac{(E-E_0e^{-ct})^2 c}{(1-e^{-2ct})2D}\right]} 
\end{equation}
which integrates out surprisingly beautifully to give simply
\begin{equation}
C(t)=e^{-ct}, \qquad c = \frac{\delta_E^2}{2\sigma^2\tau}.
\end{equation}
We see
that all information on the diffusion rate $D$ has been removed from
the correlation function, only the relation between step length and
the standard deviation remains.

\section{Statistics of steps at low temperatures}
\label{stepsappendix}
We made a simple analysis to see whether any information could be
extracted from the distribution of the steps observed. We define a
'step' by the setting of a new record low energy. We define $t_w$ to
be the waiting time from one record energy to the next, while the
energy difference between the two records is defined as the step size
$\delta_s$. If multiple consecutive jumps take the system down in
energy, only the state after the last one is accepted as a 'step'.
This still gives some artificial small $\delta_s$ whenever it takes
several steps before a local minimum is reached, so we can ignore
those steps that correspond to very small $\delta_s$. In this region
there is also a rounding error in the saved files, giving an
artificial discretization of measured $\delta_s$.

There is also evidence of steps going up in energy. These will not be
found by our algorithm, but for sufficiently low temperatures they do
not seem to play an important role.

\begin{figure}
\begin{tabular}{cc}
a) & b)\\
\includegraphics[width=7cm]{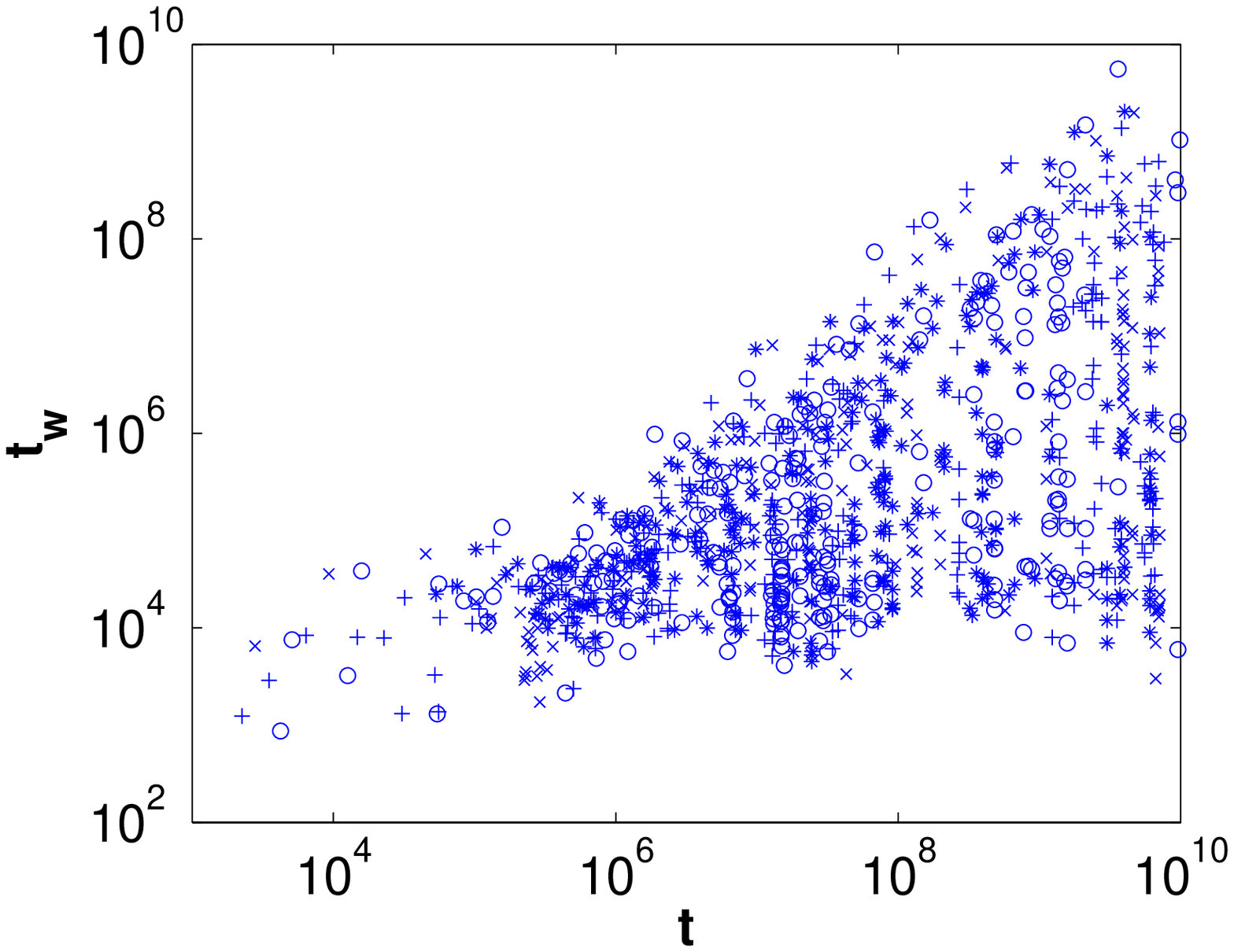}&
\includegraphics[width=7cm]{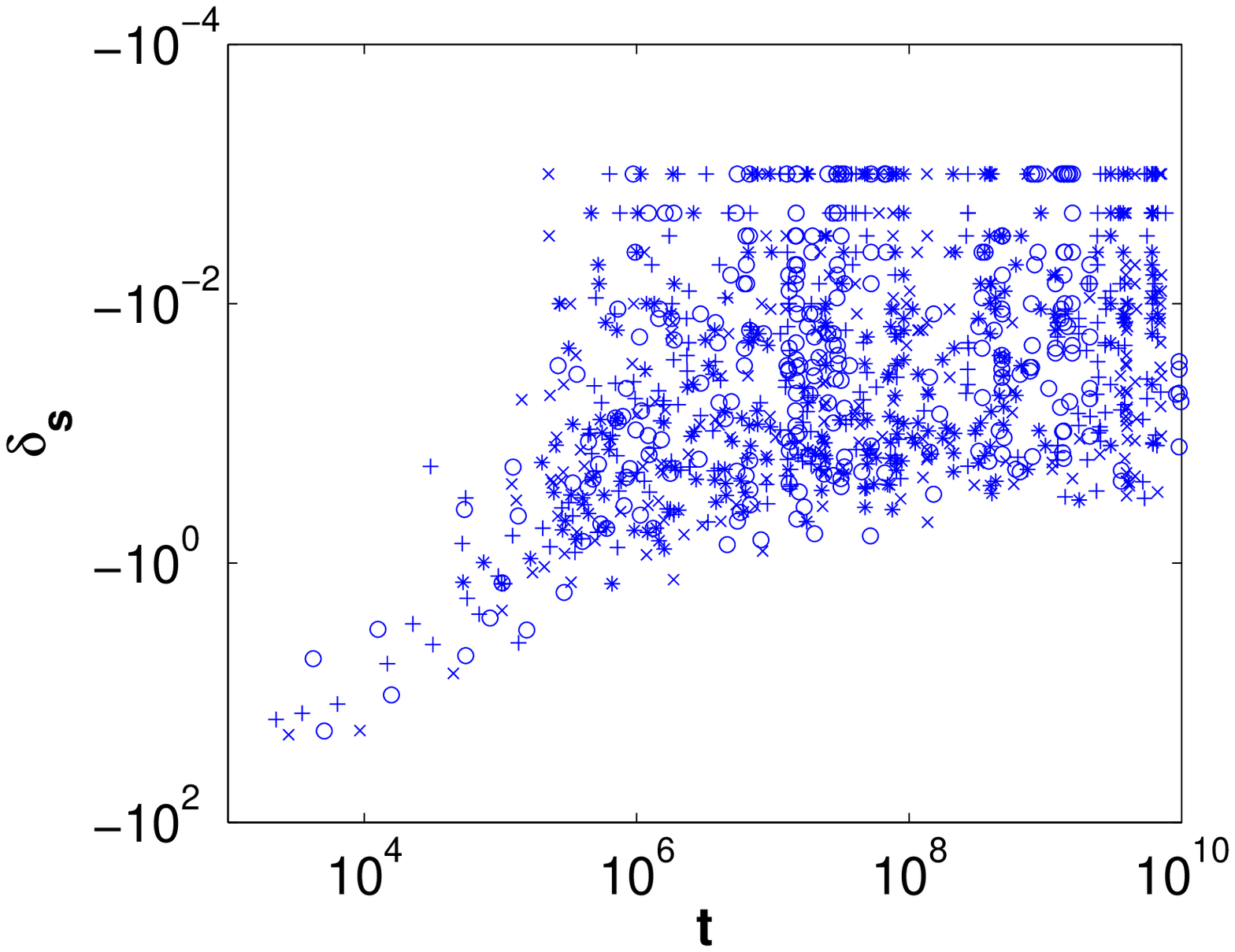}
\end{tabular}
\caption{\label{steps}a) $t_w$ vs $t$ b)$\delta_s$ vs $t$, four separate runs (o,+,*,x) at $T=0.001$.}
\end{figure}

Plots of $t_w$ and $\delta_s$ as functions of time are given in Fig.
\ref{steps}. The data are for four different relaxations at $T=0.001$.
We see that while there is a tendency that the step size $|\delta_s|$
decreases with time, the $t_w$ shows a much more pronounced behavior.
The maximum waiting time increases close to linearly with $t$. Thus
the slowing down of the relaxation seems to be due to longer waiting
times than due to smaller records being set. We have not attempted any
theoretical explanation of these results.


\begin{thebibliography}{99}
\bibitem{Mott} N. F. Mott, J. Non-Cryst. Solids {\bf 1}, 1 (1968).
\bibitem{Davies} J. H. Davies, P. A. Lee and T. M. Rice, Phys. Rev.
  Letters, {\bf 49}, 758 (1982).
\bibitem{EfrosShklovskii} B. I. Shklovskii and A. L. Efros, Electron Properties of Doped Semiconductors, Springer Series in Solid-State Sciences {\bf 45} (Springer, Berlin, 1984)\\A. L. Efros and B. I. Shklovskii, J. Phys. C {\bf 8}, 49 (1975).
\bibitem{Ovadyahu} Z. Ovadyahu, Phys. Rev. B. {\bf 73}, 214208 (2006).
\bibitem{Kar}S. Kar, A. K. Raychaudhuri, A. Ghosh, H. v. L\"ohneysen and G. Weiss, Phys. Rev. Letters{\bf 91} 216603 (2003).
\bibitem{Kurosawa} T. Kurosawa and H. Sugimoto, Progr. Theor. Pys. Suppl. {\bf 57} 217 (1975).
\bibitem{MATULEWSKI} J. Matulewski, S. D. Baranovskii and P. Thomas, 
Physica status solidi. B. {\bf 245} 3 pp. 481-484 (2008).
\bibitem{DiazSanchez} A. D\'iaz-S\'anchez, A. M\"obius, M. Ortu\~no, A. Neklioudov and M. Schreiber, Phys. Rev. B {\bf 62} 8030 (2000)
\bibitem{Pollak} A. M. Somoza, M. Ortu\~no, M. Pollak, Phys. Rev. B {\bf 73}, 045123 (2006)
\bibitem{Metropolis} N. Metropolis, A. W. Rosenbluth, M. N. Rosenbluth, A. H. Teller, E. Teller, J. Chem. Phys. {\bf 21} 1087 (1953)
\bibitem{Tsigankov} 
D. N. Tsigankov, E. Pazy, B. D. Laikhtman, and A. L. Efros, Phys. Rev. B, {\bf 68} 184205 (2003)
\bibitem{Kolton} A. B. Kolton, D. R. Grempel, D. Dom\'inguez, Phys. Rev. B {\bf 71} 024206 (2005)
\bibitem{GrannanYu} E. R. Grannan and C. C. Yu, Phys. Rev. Letters, {\bf 71} 3335 (1993)
\bibitem{MobiusThomas} A. M\"obius, P. Thomas, Phys. Rev. B {\bf 55} 7460 (1997)\\ A. M\"obius, P. Thomas, J. Talamantes and C. J. Adkins, Phil. Mag. B {\bf 81}, 1105 (2001)
\bibitem{Ortuno} M. Ortu\~no, M. Caravaca, A. M. Somoza, Phys. Stat. Sol. (c) {\bf 5}, 3 647-679 (2008)
\bibitem{Somoza} A. M. Somoza, M. Ortu\~no, M. Caravaca and M. Pollak, Phys. Rev. Letters {\bf 101} 056601 (2008)
\bibitem{Ogielski} A. T. Ogielski, Phys. Rev. B, {\bf 32} 7384 (1985)
\bibitem{MobiusRichterDrittler} A. M\"obius, M. Richter and B. Drittler, Phys. Rev. B {\bf 45} 11568 (1992)
\bibitem{Glatz} A. Glatz, V. M. Vinokur, J. Bergli, M. Kirkengen, Y. M. Galperin, Journ. Stat. Mech. (2008) P06006
\bibitem{Pikus} F. G. Pikus and A. L. Efros, Phys. Rev. Lett. {\bf 73}, 3014 - 3017 (1994)
\bibitem{Grempel} D. R. Grempel, Europhys. Letters, {\bf 66} 6 845-860 (2004)
\bibitem{Jund} P. Jund, R. Jullien, I. Campbell, Phys. Rev. E {\bf 63}, 036131 (2001)
\bibitem{Campbell} I. A. Campbell, J. Phys. (France) Lett. {\bf 46}, L1159 (1985)\\
I. A. Campbell, Phys. Rev. B {\bf 33} 3587 (1986)\\
I. A. Campbell, J. M. Flesselles, R. Jullien and R. Botet, J. Phys. C {\bf 20}, L47 (1987)\\
N. Lemke and I. A. Campbell, Physica A {\bf 230}, 554 (1996)
\bibitem{Tsigankov2} D. N. Tsigankov and A. L. Efros, Phys. Rev. Letters, {\bf 88} 176602 (2002)
\bibitem{Matulewski2} J. Matulewski, S. D. Baranovskii and P. Thomas, Phys. Stat. Sol. (c), Volume 5, Issue 3 (p 694-698) (2008)
\bibitem{Chandrasekar} S. Chandrasekar, Reviews of Modern Physics {\bf 15} 1 (1943)
\end{thebibliography}
\end{document}